\documentclass[12pt]{article}
\usepackage{amsmath}
\usepackage{amssymb}
\usepackage{graphicx}
\usepackage{subcaption}
\usepackage{url}
\usepackage{comment}
\usepackage{listings}
\usepackage[sort&compress, numbers, merge]{natbib}

\usepackage[utf8]{inputenc}

\usepackage[colorlinks=true, linkcolor=blue, citecolor=blue,
urlcolor=black]{hyperref} 

\title{WIMP Dark Matter Abundance and Standard Model Thermodynamics}
                 
\begin{document}
\baselineskip 0.6cm

\def\simgt{\mathrel{\lower2.5pt\vbox{\lineskip=0pt\baselineskip=0pt
           \hbox{$>$}\hbox{$\sim$}}}}
\def\simlt{\mathrel{\lower2.5pt\vbox{\lineskip=0pt\baselineskip=0pt
           \hbox{$<$}\hbox{$\sim$}}}}
\def\simprop{\mathrel{\lower3.0pt\vbox{\lineskip=1.0pt\baselineskip=0pt
             \hbox{$\propto$}\hbox{$\sim$}}}}
\def\tr{\mathop{\rm tr}}
\def\SU{\mathop{\rm SU}}

\def\thefootnote{\fnsymbol{footnote}}

\begin{titlepage}

\begin{flushright}
IPMU20-0047\\
KANAZAWA-20-03\\
MPP-2020-62

\end{flushright}

\vskip 1.1cm

\begin{center}

{\Large \bf 
Precise WIMP Dark Matter Abundance and Standard Model Thermodynamics
}

\vskip 1.2cm
Ken'ichi Saikawa$^{1\footnotemark,2}$ and Satoshi Shirai$^3$
\vskip 0.5cm

{\it
$^1${
Institute for Theoretical Physics, Kanazawa University,\\
Kakuma-machi, Kanazawa, Ishikawa 920-1192, Japan
}
\\
$^2${
Max-Planck-Institut f\"ur Physik (Werner-Heisenberg-Institut),
\\F\"ohringer Ring 6, D-80805 M\"unchen, Germany
}
\\
$^3${Kavli Institute for the Physics and Mathematics of the Universe (WPI), \\The University of Tokyo Institutes for Advanced Study, The University of Tokyo, Kashiwa
 277-8583, Japan}
}

\vskip 1.0cm

\abstract{
A weakly interacting massive particle (WIMP) is a leading candidate of the dark matter.
The WIMP dark matter abundance is determined by the freeze-out mechanism.
Once we know the property of the WIMP particle such as the mass and interaction, we can predict the dark matter abundance.
There are, however, several uncertainties in the estimation of the WIMP dark matter abundance.
In this work, we focus on the effect from Standard Model thermodynamics.
We revisit the estimation of the WIMP dark matter abundance and its uncertainty due to the equation of state (EOS) in the Standard Model.
We adopt the up-to-date estimate of the EOS of the Standard Model in the early Universe and find nearly 10\% difference in the 1-1000 GeV dark matter abundance, compared to the conventional estimate of the EOS.
}

\end{center}

\footnotetext{Current affiliation.}

\end{titlepage}

\renewcommand{\thefootnote}{\arabic{footnote}}
\setcounter{footnote}{0}

\section{Introduction}
\label{sec:intro}
The thermodynamics in the Standard Model (SM) plays a crucial role in the evolution of the Universe.
Any cosmological objects generated in the early Universe, such as baryon, dark matter (DM), radiation, and gravitational wave, are either directly or indirectly affected by the SM thermodynamics and evolved to the present time.
Now, the cosmological parameters are very precisely measured \cite{Aghanim:2018eyx}.
When we test a theory with the observed cosmological parameters, we need to care about the SM thermodynamics to get precise theory prediction.

In this paper, we revisit the impact of the SM thermodynamics on the abundance of a weakly interacting massive particle (WIMP) DM.
The WIMP with a mass between $\mathcal{O}(1)$\,MeV\,\cite{Boehm:2002yz,Boehm:2003bt,Serpico:2004nm,Steigman:2013yua,Nollett:2013pwa,Escudero:2018mvt,Depta:2019lbe,Sabti:2019mhn} -- $\mathcal{O}(100)$\,TeV\,\cite{Griest:1989wd, Hamaguchi:2007rb, *Hamaguchi:2008rv, *Hamaguchi:2009db, Murayama:2009nj, Hambye:2009fg, Antipin:2014qva, Antipin:2015xia, Gross:2018zha, Fukuda:2018ufg} is the most promising candidate of the DM and its cosmic abundance is determined by the freeze-out mechanism \cite{Bernstein:1985th,Srednicki:1988ce}.
Many popular extensions of the Standard Model predict such a WIMP candidate.

In the freeze-out mechanism, the DM abundance is mainly determined by the DM annihilation cross section in the early Universe.
An important feature of the freeze-out mechanism is that the DM abundance can be estimated with the low-energy feature of the WIMP and independent of the initial condition of the Universe.
Therefore if we can measure the particle property of the WIMP, such as the mass and interaction strength, we can reconstruct the WIMP DM abundance.
The collider and (in)direct DM detection experiments can measure these properties and a successful reconstruction of the DM abundance is a crucial test of the WIMP paradigm.

For instance, in the case of electroweakly interacting fermion DMs \cite{Cirelli:2005uq,Cirelli:2007xd,Cirelli:2009uv,Nagata:2014aoa}, such as a Higgsino and wino, the WIMP relic density depends only on the DM mass, and  
future lepton colliders can measure the DM mass with accuracy of $\mathcal{O}(1)$\% \cite{Berggren:2013vfa,Baer:2014yta,Fujii:2017ekh}.
This fact might imply that one can predict the DM abundance with
$\mathcal{O}(1)$\% accuracy, in principle.
For other DM candidates, there are studies for determination of the DM property at colliders \cite{Polesello:2004qy, Battaglia:2004mp, Weiglein:2004hn,Baltz:2006fm, Asakawa:2009qb}.

However, even if the low-energy DM parameters are well measured, there still remain several theoretical uncertainties 
in the DM abundance estimation.
An important uncertainty is the DM annihilation rate in the early Universe.
In particular, the DM with long range forces, non-perturbative effects, such as Sommerfeld enhancement \cite{Hisano:2003ec,Hisano:2006nn} and bound state formation \cite{Pospelov:2008jd,MarchRussell:2008tu,Shepherd:2009sa,Feng:2009mn,vonHarling:2014kha} have significant impacts on the DM annihilation and introduce sizable theoretical uncertainties. 
An improvement of these calculations is necessary for more precise estimation of the DM abundance, and these effects are intensively studied \cite{Berger:2008ti,Blum:2016nrz,Mitridate:2017izz,Baldes:2017gzw,*Harz:2017dlj,*Harz:2018csl,*Harz:2019rro,*Oncala:2019yvj,Biondini:2017ufr,*Biondini:2018pwp,*Biondini:2018xor,*Biondini:2018ovz,Binder:2018znk,Binder:2019erp}.

Another uncertainty comes from the SM thermodynamics.
The equation of state (EOS) of the SM plasma plays an important role in the expansion of the Universe and affects the DM abundance estimation.
In the SM sector, there are several non-trivial ingredients such as the QCD crossover and electoweak crossover, which lead to relatively large uncertainty in the EOS estimation.

In this paper, we revisit the estimation of the WIMP relic abundance in light of the thermodynamics in the SM.
Recently, we studied the thermodynamics of the SM in the context of the gravitational wave spectrum computation, 
by adopting the results of up-to-date lattice and perturbative  calculations \cite{Saikawa:2018rcs}.
We apply this estimate to the DM freeze-out scenario and discuss the uncertainty of the DM abundance estimation originated from the SM thermodynamics.

\section{Effective Degrees of Freedom in the Standard Model}
\label{sec:dof_sm}

In cosmology, we often use the effective relativistic degrees of freedom to estimate the abundance of various relics.
These quantities are defined as follows:
\begin{align}
g_{\rho}(T) \equiv \frac{\rho(T)}{\left[\frac{\pi^2T^4}{30}\right]},\quad g_s(T) \equiv \frac{s(T)}{\left[\frac{2\pi^2T^3}{45}\right]},
\label{geff_definitions}
\end{align}
where $\rho(T)$ and $s(T)$ are the energy density and entropy density of the SM plasma at temperature $T$.
Using thermodynamic equations,
\begin{align}
\rho(T) = T\frac{dp}{dT}(T) - p(T),\quad s(T) = \frac{\rho(T)+p(T)}{T},
\end{align}
we can rewrite Eq.~\eqref{geff_definitions} in terms of the pressure $p(T)$ and its derivative with respect to $T$:
\begin{align}
g_{\rho}(T) = \frac{30}{\pi^2}\left[\Delta(T) + \frac{3p(T)}{T^4}\right],\quad 
g_s(T) = \frac{45}{2\pi^2}\left[\Delta(T) + \frac{4p(T)}{T^4}\right],
\end{align}
where
\begin{align}
\Delta(T) \equiv \frac{\rho(T)-3p(T)}{T^4} = T\frac{d}{dT}\left\{\frac{p(T)}{T^4}\right\}
\end{align}
is called the trace anomaly. This quantity can be related to the EOS of the SM plasma,
\begin{align}
w(T) \equiv \frac{p(T)}{\rho(T)} = \frac{1}{3} - \frac{\Delta(T)}{\left[\frac{\pi^2 g_{\rho}(T)}{10}\right]}.
\end{align}
Note that the EOS parameter becomes exactly $w=1/3$ if $\Delta(T)=0$.
On the other hand, if $\Delta(T)$ becomes nonzero we expect that the value of the EOS parameter
deviates from $1/3$ and that there are some changes in $g_{\rho}(T)$ and $g_s(T)$.
Such a deviation is caused not only by the change in the relativistic degrees of freedom but also by the effect of particle interactions.

In order to estimate the effective degrees of freedom precisely, 
it is important to analyze the thermodynamic quantities including the effect of interactions of elementary particles.
Recently, there have been a lot of developments in the calculation of thermodynamic quantities in the SM including 
the analysis of the neutrino decoupling, QCD crossover, and electroweak crossover.
Such results were collected and combined to estimate the effective degrees of freedom in the SM and their uncertainty at arbitrary temperatures in Ref.~\cite{Saikawa:2018rcs}.
In the following, we briefly review these previous findings.

At temperatures below a few MeV, the effective degrees of freedom can be evaluated quite accurately 
thanks to the detailed calculations of the neutrino decoupling in Refs.~\cite{Mangano:2001iu,Mangano:2005cc,deSalas:2016ztq}.
The only concern is the effect of muons, which were neglected in those references.
In Ref.~\cite{Saikawa:2018rcs}, we found that the asymptotic value of $g_s$ at low temperatures is affected by at most $0.1\%$ when muons are included.
Regarding this fact, we assign the uncertainty of $0.1\%$ to $g_{\rho}(T)$ and $g_s(T)$ at temperatures below $10\,\mathrm{MeV}$.

In Ref.~\cite{Saikawa:2018rcs}, it was also pointed out that the value of $g_s$ after the neutrino decoupling becomes
\begin{align}
g_{s,0} \simeq 3.931 \pm 0.004,
\label{gs0_value}
\end{align}
which is slightly larger than the canonical value, $g_{s,0} = 3.91$ [e.g. Ref.~\cite{Kolb:1990vq}], which adopts no loop correction and sudden decouple approximation.\footnote{The result shown in Eq.~\eqref{gs0_value}
agrees with a value reported recently in Ref.~\cite{Escudero:2020dfa} based on a simplified method to analyze the neutrino decoupling.}
This modification is due to the fact that we take account of the residual interaction of neutrinos with electrons and positrons in the cosmic plasma, and that we include leading order quantum electrodynamics (QED) corrections on 
the thermodynamic quantities of the electromagnetic plasma.
Note that the value of $g_{s,0}$ affects the estimate of the DM abundance [see Eq.~\eqref{eq:Omega_DM_analytical}].
Therefore, there can be about $0.5\,\%$ change of the DM abundance according to the precise estimate of $g_{s,0}$.

At temperatures across $T\sim \mathcal{O}(100)\,\mathrm{MeV}$, we must take account of the effect of the QCD crossover.
The contribution of strongly interacting particles to thermodynamic quantities across the QCD crossover
can be evaluated by considering three regimes:
the hadronic regime ($T \lesssim \mathcal{O}(100)\,\mathrm{MeV}$), non-perturbative QCD regime ($T \gtrsim \mathcal{O}(100)\,\mathrm{MeV}$),
and perturbative QCD regime ($T \gg \mathcal{O}(100)\,\mathrm{MeV}$).
In the low temperature hadronic regime, the system might be approximated by free hadrons and resonances (hadron resonance gas model),
which can be compared with the results of lattice QCD simulations at the non-perturbative regime~\cite{Borsanyi:2010cj}.
The state-of-the-art results of $2+1+1$ flavor lattice QCD were presented in Ref.~\cite{Borsanyi:2016ksw},
which are consistent with the estimate based on the hadron resonance gas model at $T \lesssim 100\,\mathrm{MeV}$ within the range of uncertainty.
The uncertainty of the lattice data at $T\sim 100\,\mathrm{MeV}$ amounts to $\lesssim 13\%$ errors in $g_{\rho}(T)$ and $g_s(T)$ at that temperature~\cite{Saikawa:2018rcs}.

At sufficiently high temperatures, we can use the perturbative method to estimate the thermodynamic quantities for QCD.
In the literature, the pressure for QCD was calculated up to $\mathcal{O}(g^6)$ of weak coupling expansion~\cite{Kajantie:2002wa},
but the perturbative results show a poor convergence at $T \lesssim \mathcal{O}(10)\,\mathrm{GeV}$.
Hence it is necessary to perform an interpolation between the lattice QCD results at low temperatures and 
perturbative results at high temperatures.
It turned out that the final result is sensitive to the interpolation procedure,
and this amounts to at most $9\%$ uncertainty in $g_{\rho}(T)$ and $g_s(T)$ at $T = \mathcal{O}(1 \textendash 10)\,\mathrm{GeV}$~\cite{Saikawa:2018rcs}.

In addition to the contribution from QCD, we also have to take account of the effect of the electroweak crossover.
Again this can be addressed by considering three different steps.
First, at temperatures lower than the critical temperature of the electroweak crossover, 
the thermodynamic quantities can be evaluated by using the one-loop~\cite{Laine:2006cp} or 
two-loop~\cite{Farakos:1994kx,Farakos:1994xh,Kajantie:1995dw,Kajantie:1995kf} Coleman-Weinberg potential.
Second, at sufficiently high temperatures, we rely on the perturbative results evaluated up to $\mathcal{O}(g^5)$~\cite{Gynther:2005dj,Laine:2015kra}.
Finally, at the intermediate temperatures including the critical temperature of the electroweak crossover, 
we can use the results of lattice simulations~\cite{DOnofrio:2015gop}.
In Ref.~\cite{Saikawa:2018rcs}, we observed that the uncertainty induced by the interpolation of these results is less significant
compared to that arising from the QCD sector.

The full results of the effective degrees of freedom $g_{\rho}(T)$ and $g_s(T)$ and their uncertainty at arbitrary temperatures became available 
as tabulated data~\cite{Saikawa:2018rcs}, which we use in the next section to calculate the relic DM abundance.
Here we emphasize that the errors in $g_{\rho}(T)$ and $g_s(T)$ shown in those data are not statistical, but
rather correspond to a typical range obtained by various possible computational methods or interpolation procedures.
As the dominant source of these uncertainties is the estimation of the thermodynamic quantities in QCD at $T = \mathcal{O}(1 \textendash 10)\,\mathrm{GeV}$,
the resolution of them would require some further non-perturbative analysis at high temperatures.

\begin{figure}[tbp]
	\centering
	\subcaptionbox{\label{fig:gs}$g_s(T)$ }{\includegraphics[width=0.47\textwidth]{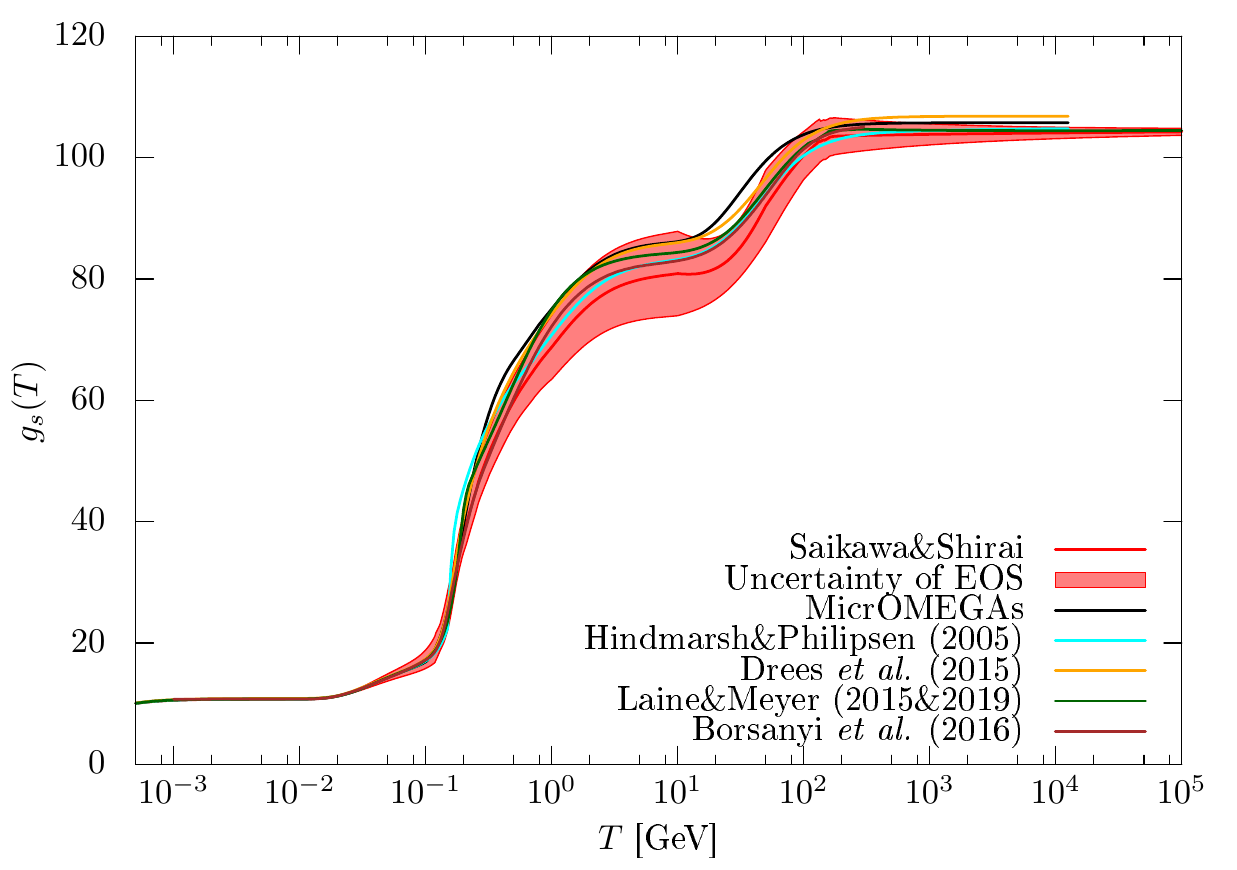}}
	\subcaptionbox{\label{fig:gslalf}$g_*^{1/2}(T)$}{\includegraphics[width=0.47\textwidth]{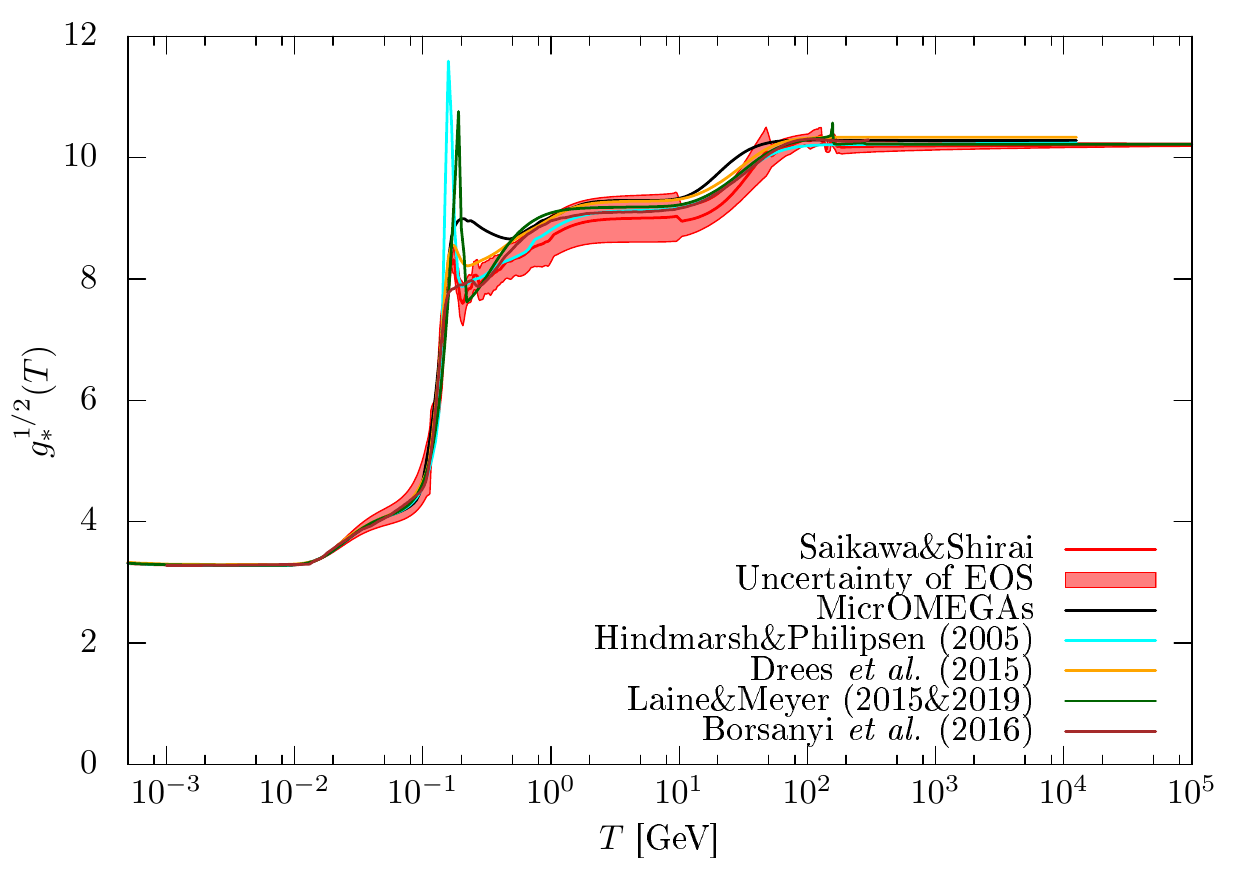}}
\caption{
Temperature dependence of the effective degrees of freedom for the entropy density $g_s$ (left panel) and $g_*^{1/2}$ defined in Eq.~\eqref{eq:gshalf} (right panel).
The red shaded region in the left panel shows the uncertainty from the thermodynamics of the Standard Model obtained in Ref.~\cite{Saikawa:2018rcs}.
Results of several previous papers are also shown.
In the right panel, the temperature derivative for the data from Ref.~\cite{Saikawa:2018rcs} is computed by using their median.
}
\label{fig:eos}
\end{figure}

In Fig.~\ref{fig:eos}, we show the result of $g_s(T)$ obtained in Ref.~\cite{Saikawa:2018rcs} and $g_*^{1/2}(T)$, which is often used for the WIMP relic abundance estimation and defined as 
\begin{align}
g_*^{1/2}(T) \equiv 
\frac{g_{s} (T)}{g_{\rho}^{1/2}(T)} \left(
1 + \frac{T}{3 g_s(T)} \frac{d  g_{s}(T)}{dT}\right),
\label{eq:gshalf}
\end{align}
and compare them to previous results. 
The data of the effective degrees of freedom used in {\tt micrOMEGAs}~\cite{Belanger:2018mqt} by default are based on Refs.~\cite{Srednicki:1988ce,Gondolo:1990dk},
where the contribution of strongly interacting particles is included by considering the gas of free hadrons at low temperatures and that of free quarks and gluons at high temperatures,
and by interpolating them at $T=150\,\mathrm{MeV}$ via a smooth function. This interpolation function was introduced just to make the transition smooth and not based on dynamical considerations.
Hindmarsh and Philipsen (2005)~\cite{Hindmarsh:2005ix} introduced the lattice results from Ref.~\cite{Karsch:2000ps} to describe the effective degrees of freedom around the QCD transition.
In Fig.~\ref{fig:eos}, we plot the ``EOS B" model in their paper. In this model there is an artificial sharp switch from the hadronic gas to the lattice data at $T=154\,\mathrm{MeV}$,
which leads to a sharp peak in the plot of $g_*^{1/2}$.
Drees {\it et al.} (2015)~\cite{Drees:2015exa} used the results of $2+1$ flavor lattice QCD from Ref.~\cite{Bazavov:2014pvz}. 
While the estimation of the QCD contribution was improved in their approach, Higgs and electroweak gauge bosons were treated as free massive particles without including the effect of their interactions.
The corrections due to electroweak and Yukawa interactions in the SM were introduced by Laine and Meyer (2015)~\cite{Laine:2015kra},
but their result is based on the lattice data~\cite{Boyd:1996bx} which do not include dynamical quarks.
Borsanyi {\it et al.} (2016)~\cite{Borsanyi:2016ksw} introduced the state-of-the-art results of $2+1+1$ flavor lattice QCD and also included the electroweak corrections obtained in Ref.~\cite{Laine:2015kra}.
Note that, depending on the treatment of the lattice data and perturbative calculation, the estimated effective degrees of freedom can vary by about 10\% as seen in  Fig.~\ref{fig:eos}.

Our previous result~\cite{Saikawa:2018rcs} is basically consistent with Borsanyi {\it et al.} (2016) as we used the same lattice data and took a similar approach to describe the electroweak corrections,
but we assigned a more conservative estimate for the uncertainty of the effective degrees of freedom as described in this section and shown as the red shaded region in Fig.~\ref{fig:eos}.

\section{Dark Matter Abundance}
\label{sec:DM_abundance}
In this section, we discuss the impact of the EOS reviewed in the previous section on the WIMP relic density.
The evolution of the DM number density $n_{\mathrm{DM}}$ is given by the following Boltzmann equation:

\begin{align}
\frac{d n_{\mathrm{DM}}}{d t} + 3 H n_{\mathrm{DM}}
 = -\langle \sigma v \rangle (n_{\mathrm{DM}}^2-n_{\mathrm{DM, eq}}^2),
 \label{eq:Boltzmann}
\end{align}
where $H$ is the Hubble parameter, $ \langle \sigma v \rangle$  thermal average of the annihilation cross section times relative velocity between the two annihilating DM, $n_{\mathrm {DM, eq}}$ is the DM number density in thermal equilibrium.\footnote{If $m_{\rm DM} \lesssim 10$ MeV, 
the DM annihilation after the neutrino decoupling modifies the temperatures of the neutrinos and photons.
This leads to the shift of the effective neutrino number $\Delta N_{\rm eff}$ and affects the DM abundance estimation \cite{Steigman:2013yua,Nollett:2013pwa,Escudero:2018mvt,Sabti:2019mhn}.
In this work, we neglect this effect for simplicity.}
Equivalently, it is often convenient to use the following form of the Boltzmann equation in terms of the DM yield $Y_{\mathrm{DM}} \equiv  n_{\mathrm{DM}}/s$ and mass to temperature ratio $x\equiv m_{\mathrm{DM}}/T$:
\begin{align}
\frac{d Y_{\mathrm {DM}}}{d x}
 =- 
\left(\frac{45}{8\pi^2 } \right)^{-1/2}
\frac{g_*^{1/2} m_{\mathrm{DM}} M_{\mathrm {P}} }{x^2}
 \langle \sigma v \rangle (Y_{\mathrm{DM}}^2-Y_{\mathrm{DM, eq}}^2
 ),
 \label{eq:Boltzmann_yield}
\end{align}
where $ M_{\mathrm {P}} = 2.4353 \times 10^{18}$ GeV is the reduced Planck mass and $g_*^{1/2}$ is given in Eq.~\eqref{eq:gshalf}.

The DM abundance is approximately given by:
\begin{align}
\Omega_{\mathrm{DM}} h^2 \sim 
\frac{s_0 h^2}{\rho_{\mathrm {c}}}
\left(\frac{45}{8\pi^2 {g}_{\rho}(T_{\mathrm{FO}} ) } \right)^{1/2}
\frac{m_{\mathrm{DM}} }{T_{\mathrm{FO}}  M_{\mathrm {P}}  \langle \sigma v \rangle  },
\label{eq:Omega_DM_analytical}
\end{align}
where $h$ is the renormalized Hubble parameter, $H_0 = 100\,h\,\mathrm{km}\,\mathrm{sec}^{-1}\mathrm{Mpc}^{-1}$,
$s_0$ is the entropy density after the neutrino decoupling,
$\rho_{\mathrm {c}}$ is the critical density at the present time,
and $T_{\mathrm{FO}}$ is the  freeze-out temperature and approximately given as~\cite{Kolb:1990vq}
\begin{align}
\frac{m_{\mathrm {DM}}}{T_{\mathrm{FO}}} \sim \log(0.2 M_{\mathrm {P}} g_{\rm DM} {g}_{\rho}^{-1/2}   m_{\rm DM}\langle \sigma v \rangle ) \sim 10-30.
\end{align}
Here $g_{\rm DM}$ is the internal degrees of freedom of the WIMP particle.
From the rough estimate of the abundance \eqref{eq:Omega_DM_analytical}, we see that the DM abundance is approximately proportional to $(g_{\rho}(m_{\rm DM}/20))^{-1/2}$.

\subsection*{$s$- and $p$-wave annihilating DM }
The annihilation cross section $\langle \sigma v \rangle$ depends on the details of DM models.
Here let us discuss simplified DM annihilation cases.
In the following, we assume the DM is always in kinetic equilibrium with the SM or photon, for simplicity.
In the non-relativistic limit $v \ll 1$, the cross section may be expanded as 
\begin{align}
    \sigma v \simeq a_s  + b_p v^2+\cdots. \label{eq:partial}
\end{align}
By taking the thermal average of the annihilation rate, we have
\begin{align}
    \langle \sigma v  \rangle \simeq a_s  + b_p \frac{6 T}{m_{\rm DM}}+\cdots.
\end{align}
To see the impact of the EOS, we consider $s$-wave ($a_s > 0$ and others zero) and $p$-wave ($b_p>0$ and others zero) annihilating WIMP.
We solve the Boltzmann equation \eqref{eq:Boltzmann} with $s$- and $p$-wave annihilation cross section up to the cosmic microwave background (CMB) era $T = 1$ eV and obtain $\Omega_{\rm DM} h^2$.
Here we assume the DM is a Majorana fermion, $g_{\rm DM}=2$.

\begin{figure}[!ht]
\centering
\includegraphics[width=0.94\textwidth]{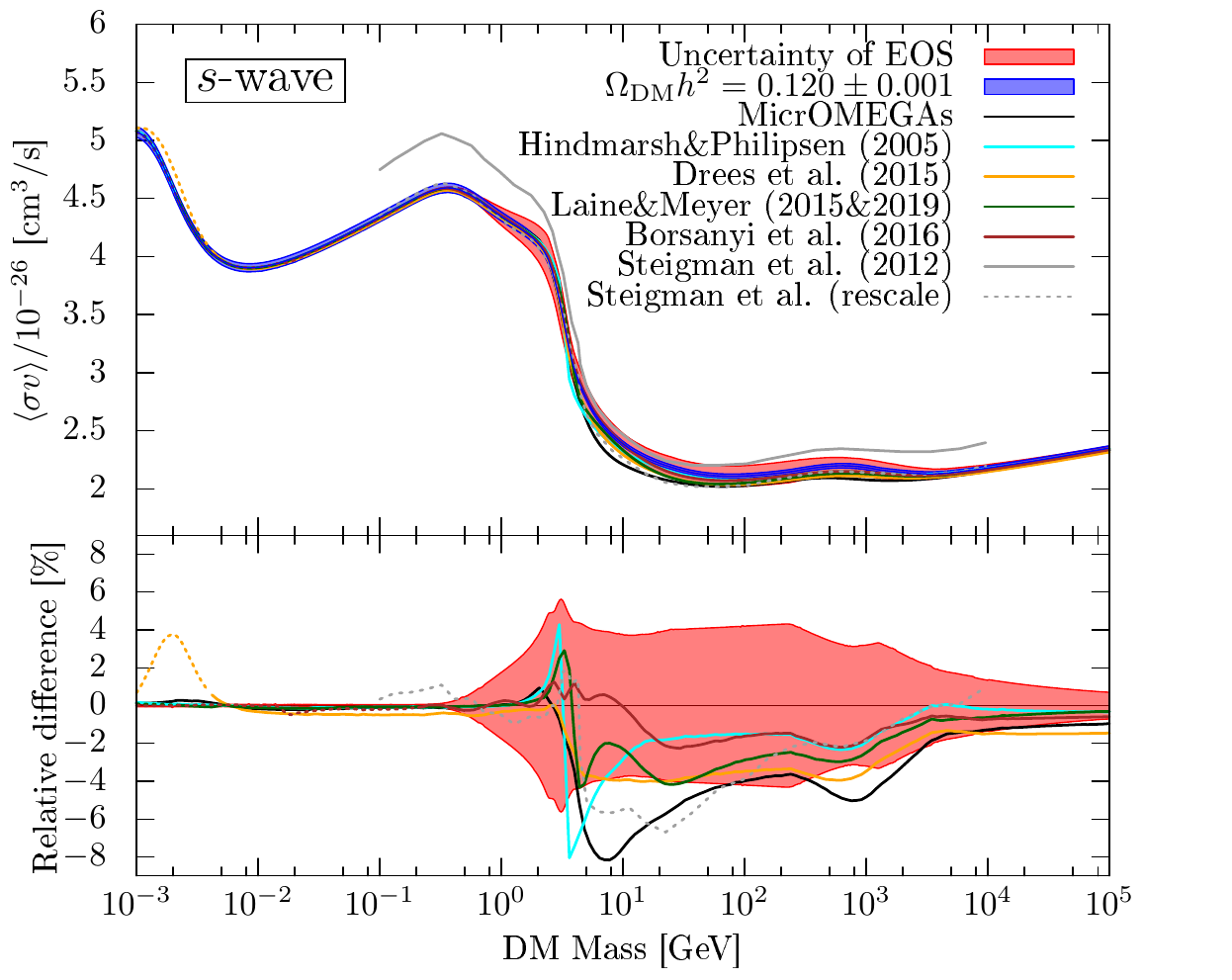}
\caption{
The annihilation cross section times velocity to realize $\Omega_{\mathrm{DM}} h^2 = 0.120 \pm 0.001$ \cite{Aghanim:2018eyx} for $s$-wave annihilating Majorana fermion DM.
The red shaded region shows the uncertainty from the thermodynamics of the SM, and the blue shows uncertainty of the DM abundance measurement.
We also show the result of Steigman {\it et al.}~\cite{Steigman:2012nb} (gray solid line), which was obtained by assuming $\Omega_{\mathrm{DM}}h^2 = 0.11$, and that multiplied by a rescaling factor 0.11/0.12 (gray dotted line) to remove the difference coming from the different assumption about $\Omega_{\mathrm{DM}}h^2$.
}
\label{fig:swave}
\end{figure}

\begin{figure}[!ht]
\centering
\includegraphics[width=0.94\textwidth]{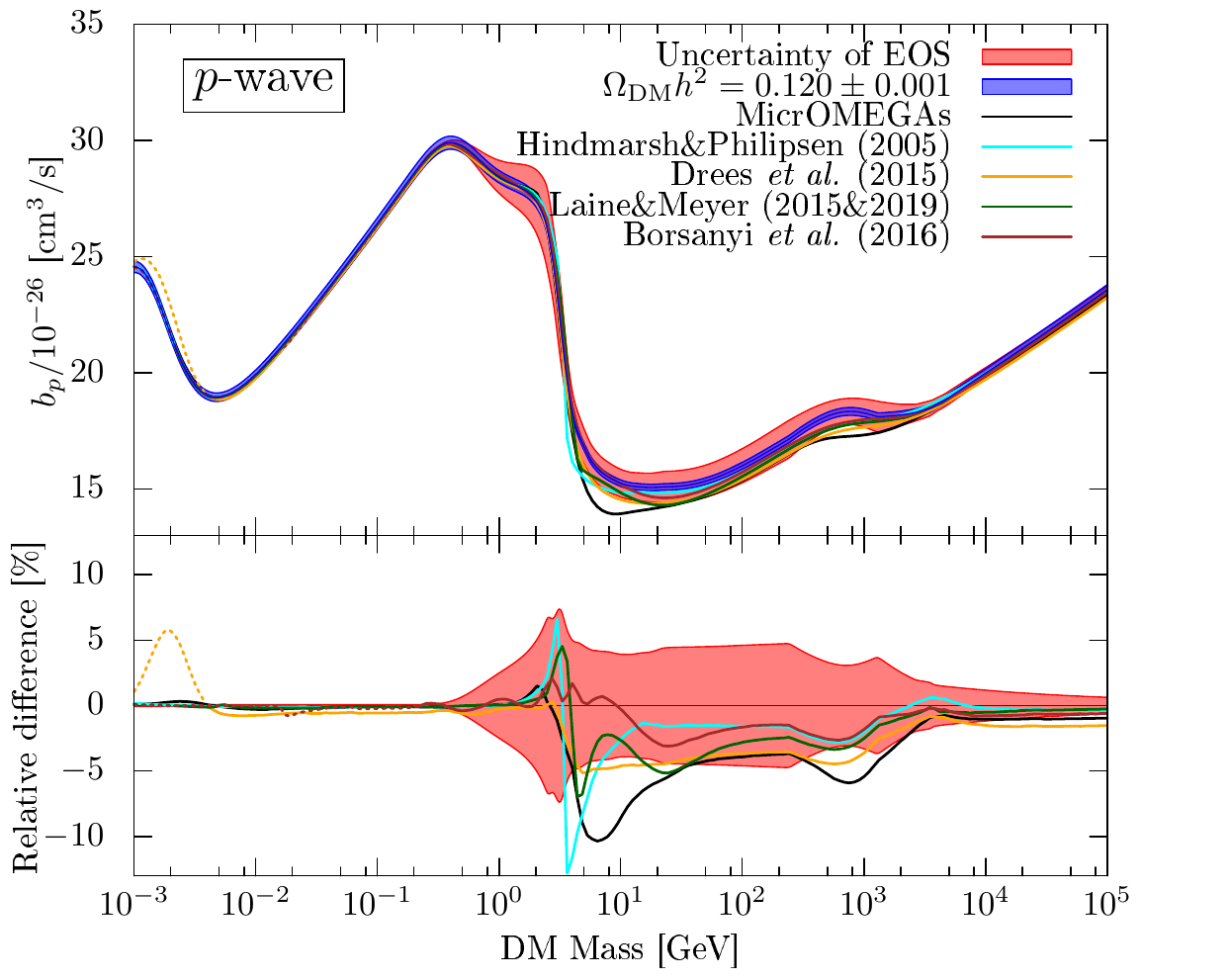}
\caption{
Same as Fig.~\ref{fig:swave} but for $p$-wave annihilating Majorana fermion DM
The definition of $b_p$ is given in Eq.~\eqref{eq:partial}.
}
\label{fig:pwave}
\end{figure}

In Figs.~\ref{fig:swave} and \ref{fig:pwave}, we show the predicted cross section to realize the current observation $\Omega_{\rm DM}h^2 = 0.120\pm 0.001$ \cite{Aghanim:2018eyx} for $s$- and $p$-wave annihilating Majorana fermion  DM, respectively.
The red shaded region shows the uncertainty from the thermodynamics of the SM, and the blue shows uncertainty of the DM abundance measurement.
We also show the predicted cross section based on other EOS estimates.
In the bottom panel, we show the relative difference between the present EOS and other EOS estimates.

For the $s$-wave annihilating DM case, we also show the result by Steigman {\it et al.}~\cite{Steigman:2012nb} for a comparison.
Our estimate is about 10\% smaller than that of Ref.~\cite{Steigman:2012nb}.
The most important difference comes from the DM abundance measured by the observation of CMB.
In our analysis, we adopt the latest Planck result $\Omega_{\mathrm{DM}}h^2 = 0.120$~\cite{Aghanim:2018eyx},
while Ref.~\cite{Steigman:2012nb} adopted $\Omega_{\mathrm{DM}}h^2 = 0.11$ based on WMAP-7~\cite{Komatsu:2010fb}.
Another important difference is the treatment of the EOS.
Reference~\cite{Steigman:2012nb} approximated $g_s = g_{\rho}$ and used the estimate of $g_{\rho}$ by Laine and Schr\"oder (2006)~\cite{Laine:2006cp}, which leads to about 10\% difference between our result and that of Ref.~\cite{Steigman:2012nb}.
In Ref.~\cite{Steigman:2012nb}, the integration of the Boltzmann equation \eqref{eq:Boltzmann_yield} ends at $x=1000$, whereas we integrate up to $T=1$ eV.
This also affects the result by a few percent.
Moreover,  the difference in $g_{s,0}$ modifies the result with sub-percent level.
When we adopt the setup of Ref.~\cite{Steigman:2012nb}, our numerical estimate agrees with the result of Ref.~\cite{Steigman:2012nb}.

In Figs.~\ref{fig:swave} and \ref{fig:pwave}, we consider a Majorana fermion DM.
We can straightforwardly apply this analysis to other types of DM, such as a real scalar DM ($g_{\rm DM}=1$) and Dirac fermion DM ($g_{\rm DM}=4$).
For the larger $g_{\rm DM}$, we need slightly larger $\langle \sigma v \rangle $ to reproduce the correct  DM abundance.
The data are available at \url{https://member.ipmu.jp/satoshi.shirai/DM2020/}.

\section{Conclusion and discussion}
\label{sec:conclusion}
In this paper, we have estimated the WIMP DM abundance based on the latest result of the SM thermodynamics.
We have found that the present result can differ from the conventional estimates by about 5-10\% for $m_{\rm DM}>1$ GeV.
We have also observed that the systematic uncertainty of the DM abundance 
coming from the estimation of the EOS in the SM is a few percent.

The largest uncertainty comes from the QCD sector.
For the temperature around ${\cal O}(0.1-1)$ GeV, the error of the EOS comes from the lattice QCD.
For the temperature greater than ${\cal O}(1)$ GeV, the lattice data is no longer available and we have to use the perturbative results.
However, the perturbative expansion is poorly convergent and we need to match the perturbative estimates to the lattice result at the low temperature.
Taking account of the uncertainty from the lattice data and that from the matching procedure,
we expect that the EOS error is $\Delta g_{\rho}/g_\rho\simeq 10$\% for $T=0.1-100$ GeV.
Therefore the uncertainty of the DM abundance for $m_{\rm DM}=1-1000$ GeV is around 5\%, as the  DM abundance is approximately proportional to $g_{\rho}(T_{\rm FO})^{-1/2}$.

Improving this uncertainty is important for the establishment of the WIMP paradigm based on the (in)direct detection and collider experiments.
In particular, future colliders can measure the DM property very well \cite{Berggren:2013vfa,Baer:2014yta,Fujii:2017ekh,Polesello:2004qy, Battaglia:2004mp, Weiglein:2004hn,Baltz:2006fm, Asakawa:2009qb}, which enables us to compute $\langle \sigma v\rangle$ precisely, predict the DM abundance and compare it with the observed abundance.
The consistent reconstruction of the DM abundance is the most striking test of the WIMP model.
For this procedure, the proper treatment of the EOS and its uncertainty are important, as the EOS uncertainty may modify the abundance estimation by 5\%-10\% as discussed in this work, and this uncertainty is significantly larger than the expected accuracy of $\langle \sigma v\rangle$ estimation at future colliders.

We distribute the data of the WIMP cross section and EOS obtained in this paper at
\url{https://member.ipmu.jp/satoshi.shirai/DM2020/}.
One can use the present EOS data in {\tt micrOMEGAs v5} \cite{Belanger:2018mqt}, by putting 
\texttt{loadHeffGeff("eos2020.dat");} in the main source code.
Note that {\tt micrOMEGAs v5} seems to adopt $g_{s,0} = 3.906$ when converting $Y_{\rm DM}$ to $\Omega_{\rm DM} h^2$, 
while in our estimation we have used $g_{s,0} = 3.931$.
Therefore, multiplying the factor $3.931/3.906 \simeq 1.006$ by the {\tt micrOMEGAs v5} output with {\tt eos2020.dat} 
reproduces our result quite accurately.
In the online distribution, we also provide the data of $\langle \sigma v\rangle$ estimated for the cases of the non-fermion WIMP DM.

\section*{Acknowledgments}
This work of SS is supported by Grant-in-Aid for Scientific Research from the Ministry of Education, Culture, Sports, Science, and Technology (MEXT), Japan, No.~17H02878, 18K13535, 19H04609 and 20H01895, and by World Premier International Research Center Initiative (WPI), MEXT, Japan.
This project has received funding from the European Union’s Horizon 2020 research and innovation programme under the Marie Sktodowska-Curie grant agreement No 690575.
KS acknowledges partial support at the Max-Planck-Institute for Physics
by the Deutsche Forschungsgemeinschaft through Grant No.\ EXC 153 (Excellence
Cluster ``Universe'') and Grant No.\ SFB 1258 (Collaborative Research
Center ``Neutrinos, Dark Matter, Messengers'') as well as by the
European Union through Grant No.\ H2020-MSCA-ITN-2015/674896 (Innovative Training Network ``Elusives'').
The work of KS at Kanazawa University is supported by Leading Initiative for Excellent Young Researchers, MEXT, Japan.

\bibliographystyle{aps}
\bibliography{ref}

\begin{thebibliography}{80}%
\makeatletter
\providecommand \@ifxundefined [1]{%
 \@ifx{#1\undefined}
}%
\providecommand \@ifnum [1]{%
 \ifnum #1\expandafter \@firstoftwo
 \else \expandafter \@secondoftwo
 \fi
}%
\providecommand \@ifx [1]{%
 \ifx #1\expandafter \@firstoftwo
 \else \expandafter \@secondoftwo
 \fi
}%
\providecommand \natexlab [1]{#1}%
\providecommand \enquote  [1]{``#1''}%
\providecommand \bibnamefont  [1]{#1}%
\providecommand \bibfnamefont [1]{#1}%
\providecommand \citenamefont [1]{#1}%
\providecommand \href@noop [0]{\@secondoftwo}%
\providecommand \href [0]{\begingroup \@sanitize@url \@href}%
\providecommand \@href[1]{\@@startlink{#1}\@@href}%
\providecommand \@@href[1]{\endgroup#1\@@endlink}%
\providecommand \@sanitize@url [0]{\catcode `\\12\catcode `\$12\catcode
  `\&12\catcode `\#12\catcode `\^12\catcode `\_12\catcode `\%12\relax}%
\providecommand \@@startlink[1]{}%
\providecommand \@@endlink[0]{}%
\providecommand \url  [0]{\begingroup\@sanitize@url \@url }%
\providecommand \@url [1]{\endgroup\@href {#1}{\urlprefix }}%
\providecommand \urlprefix  [0]{URL }%
\providecommand \Eprint [0]{\href }%
\providecommand \doibase [0]{http://dx.doi.org/}%
\providecommand \selectlanguage [0]{\@gobble}%
\providecommand \bibinfo  [0]{\@secondoftwo}%
\providecommand \bibfield  [0]{\@secondoftwo}%
\providecommand \translation [1]{[#1]}%
\providecommand \BibitemOpen [0]{}%
\providecommand \bibitemStop [0]{}%
\providecommand \bibitemNoStop [0]{.\EOS\space}%
\providecommand \EOS [0]{\spacefactor3000\relax}%
\providecommand \BibitemShut  [1]{\csname bibitem#1\endcsname}%
\let\auto@bib@innerbib\@empty
\bibitem [{\citenamefont {Aghanim}\ \emph {et~al.}(2018)\citenamefont {Aghanim}
  \emph {et~al.}}]{Aghanim:2018eyx}%
  \BibitemOpen
  \bibfield  {author} {\bibinfo {author} {\bibfnamefont {N.}~\bibnamefont
  {Aghanim}} \emph {et~al.} (\bibinfo {collaboration} {Planck}),\ }\href@noop
  {} {\  (\bibinfo {year} {2018})},\ \Eprint
  {http://arxiv.org/abs/1807.06209}{arXiv:1807.06209 [astro-ph.CO]}\BibitemShut
  {NoStop}%
\bibitem [{\citenamefont {Boehm}\ \emph
  {et~al.}(2004{\natexlab{a}})\citenamefont {Boehm}, \citenamefont {Ensslin},\
  and\ \citenamefont {Silk}}]{Boehm:2002yz}%
  \BibitemOpen
  \bibfield  {author} {\bibinfo {author} {\bibfnamefont {C.}~\bibnamefont
  {Boehm}}, \bibinfo {author} {\bibfnamefont {T.~A.}\ \bibnamefont {Ensslin}},
  \ and\ \bibinfo {author} {\bibfnamefont {J.}~\bibnamefont {Silk}},\ }\href
  {\doibase 10.1088/0954-3899/30/3/004} {\bibfield  {journal} {\bibinfo
  {journal} {J. Phys.}\ }\textbf {\bibinfo {volume} {G30}},\ \bibinfo {pages}
  {279} (\bibinfo {year} {2004}{\natexlab{a}})},\ \Eprint
  {http://arxiv.org/abs/astro-ph/0208458}{arXiv:astro-ph/0208458
  [astro-ph]}\BibitemShut {NoStop}%
\bibitem [{\citenamefont {Boehm}\ \emph
  {et~al.}(2004{\natexlab{b}})\citenamefont {Boehm}, \citenamefont {Hooper},
  \citenamefont {Silk}, \citenamefont {Casse},\ and\ \citenamefont
  {Paul}}]{Boehm:2003bt}%
  \BibitemOpen
  \bibfield  {author} {\bibinfo {author} {\bibfnamefont {C.}~\bibnamefont
  {Boehm}}, \bibinfo {author} {\bibfnamefont {D.}~\bibnamefont {Hooper}},
  \bibinfo {author} {\bibfnamefont {J.}~\bibnamefont {Silk}}, \bibinfo {author}
  {\bibfnamefont {M.}~\bibnamefont {Casse}}, \ and\ \bibinfo {author}
  {\bibfnamefont {J.}~\bibnamefont {Paul}},\ }\href {\doibase
  10.1103/PhysRevLett.92.101301} {\bibfield  {journal} {\bibinfo  {journal}
  {Phys. Rev. Lett.}\ }\textbf {\bibinfo {volume} {92}},\ \bibinfo {pages}
  {101301} (\bibinfo {year} {2004}{\natexlab{b}})},\ \Eprint
  {http://arxiv.org/abs/astro-ph/0309686}{arXiv:astro-ph/0309686
  [astro-ph]}\BibitemShut {NoStop}%
\bibitem [{\citenamefont {Serpico}\ and\ \citenamefont
  {Raffelt}(2004)}]{Serpico:2004nm}%
  \BibitemOpen
  \bibfield  {author} {\bibinfo {author} {\bibfnamefont {P.~D.}\ \bibnamefont
  {Serpico}}\ and\ \bibinfo {author} {\bibfnamefont {G.~G.}\ \bibnamefont
  {Raffelt}},\ }\href {\doibase 10.1103/PhysRevD.70.043526} {\bibfield
  {journal} {\bibinfo  {journal} {Phys. Rev. D}\ }\textbf {\bibinfo {volume}
  {70}},\ \bibinfo {pages} {043526} (\bibinfo {year} {2004})},\ \Eprint
  {http://arxiv.org/abs/astro-ph/0403417}{arXiv:astro-ph/0403417}\BibitemShut
  {NoStop}%
\bibitem [{\citenamefont {Steigman}(2013)}]{Steigman:2013yua}%
  \BibitemOpen
  \bibfield  {author} {\bibinfo {author} {\bibfnamefont {G.}~\bibnamefont
  {Steigman}},\ }\href {\doibase 10.1103/PhysRevD.87.103517} {\bibfield
  {journal} {\bibinfo  {journal} {Phys. Rev. D}\ }\textbf {\bibinfo {volume}
  {87}},\ \bibinfo {pages} {103517} (\bibinfo {year} {2013})},\ \Eprint
  {http://arxiv.org/abs/1303.0049}{arXiv:1303.0049 [astro-ph.CO]}\BibitemShut
  {NoStop}%
\bibitem [{\citenamefont {Nollett}\ and\ \citenamefont
  {Steigman}(2014)}]{Nollett:2013pwa}%
  \BibitemOpen
  \bibfield  {author} {\bibinfo {author} {\bibfnamefont {K.~M.}\ \bibnamefont
  {Nollett}}\ and\ \bibinfo {author} {\bibfnamefont {G.}~\bibnamefont
  {Steigman}},\ }\href {\doibase 10.1103/PhysRevD.89.083508} {\bibfield
  {journal} {\bibinfo  {journal} {Phys. Rev. D}\ }\textbf {\bibinfo {volume}
  {89}},\ \bibinfo {pages} {083508} (\bibinfo {year} {2014})},\ \Eprint
  {http://arxiv.org/abs/1312.5725}{arXiv:1312.5725 [astro-ph.CO]}\BibitemShut
  {NoStop}%
\bibitem [{\citenamefont {Escudero}(2019)}]{Escudero:2018mvt}%
  \BibitemOpen
  \bibfield  {author} {\bibinfo {author} {\bibfnamefont {M.}~\bibnamefont
  {Escudero}},\ }\href {\doibase 10.1088/1475-7516/2019/02/007} {\bibfield
  {journal} {\bibinfo  {journal} {JCAP}\ }\textbf {\bibinfo {volume} {1902}},\
  \bibinfo {pages} {007} (\bibinfo {year} {2019})},\ \Eprint
  {http://arxiv.org/abs/1812.05605}{arXiv:1812.05605 [hep-ph]}\BibitemShut
  {NoStop}%
\bibitem [{\citenamefont {Depta}\ \emph {et~al.}(2019)\citenamefont {Depta},
  \citenamefont {Hufnagel}, \citenamefont {Schmidt-Hoberg},\ and\ \citenamefont
  {Wild}}]{Depta:2019lbe}%
  \BibitemOpen
  \bibfield  {author} {\bibinfo {author} {\bibfnamefont {P.~F.}\ \bibnamefont
  {Depta}}, \bibinfo {author} {\bibfnamefont {M.}~\bibnamefont {Hufnagel}},
  \bibinfo {author} {\bibfnamefont {K.}~\bibnamefont {Schmidt-Hoberg}}, \ and\
  \bibinfo {author} {\bibfnamefont {S.}~\bibnamefont {Wild}},\ }\href {\doibase
  10.1088/1475-7516/2019/04/029} {\bibfield  {journal} {\bibinfo  {journal}
  {JCAP}\ }\textbf {\bibinfo {volume} {04}},\ \bibinfo {pages} {029} (\bibinfo
  {year} {2019})},\ \Eprint {http://arxiv.org/abs/1901.06944}{arXiv:1901.06944
  [hep-ph]}\BibitemShut {NoStop}%
\bibitem [{\citenamefont {Sabti}\ \emph {et~al.}(2020)\citenamefont {Sabti},
  \citenamefont {Alvey}, \citenamefont {Escudero}, \citenamefont {Fairbairn},\
  and\ \citenamefont {Blas}}]{Sabti:2019mhn}%
  \BibitemOpen
  \bibfield  {author} {\bibinfo {author} {\bibfnamefont {N.}~\bibnamefont
  {Sabti}}, \bibinfo {author} {\bibfnamefont {J.}~\bibnamefont {Alvey}},
  \bibinfo {author} {\bibfnamefont {M.}~\bibnamefont {Escudero}}, \bibinfo
  {author} {\bibfnamefont {M.}~\bibnamefont {Fairbairn}}, \ and\ \bibinfo
  {author} {\bibfnamefont {D.}~\bibnamefont {Blas}},\ }\href {\doibase
  10.1088/1475-7516/2020/01/004} {\bibfield  {journal} {\bibinfo  {journal}
  {JCAP}\ }\textbf {\bibinfo {volume} {01}},\ \bibinfo {pages} {004} (\bibinfo
  {year} {2020})},\ \Eprint {http://arxiv.org/abs/1910.01649}{arXiv:1910.01649
  [hep-ph]}\BibitemShut {NoStop}%
\bibitem [{\citenamefont {Griest}\ and\ \citenamefont
  {Kamionkowski}(1990)}]{Griest:1989wd}%
  \BibitemOpen
  \bibfield  {author} {\bibinfo {author} {\bibfnamefont {K.}~\bibnamefont
  {Griest}}\ and\ \bibinfo {author} {\bibfnamefont {M.}~\bibnamefont
  {Kamionkowski}},\ }\href {\doibase 10.1103/PhysRevLett.64.615} {\bibfield
  {journal} {\bibinfo  {journal} {Phys. Rev. Lett.}\ }\textbf {\bibinfo
  {volume} {64}},\ \bibinfo {pages} {615} (\bibinfo {year} {1990})}\BibitemShut
  {NoStop}%
\bibitem [{\citenamefont {Hamaguchi}\ \emph {et~al.}(2007)\citenamefont
  {Hamaguchi}, \citenamefont {Shirai},\ and\ \citenamefont
  {Yanagida}}]{Hamaguchi:2007rb}%
  \BibitemOpen
  \bibfield  {author} {\bibinfo {author} {\bibfnamefont {K.}~\bibnamefont
  {Hamaguchi}}, \bibinfo {author} {\bibfnamefont {S.}~\bibnamefont {Shirai}}, \
  and\ \bibinfo {author} {\bibfnamefont {T.~T.}\ \bibnamefont {Yanagida}},\
  }\href {\doibase 10.1016/j.physletb.2007.08.047} {\bibfield  {journal}
  {\bibinfo  {journal} {Phys. Lett.}\ }\textbf {\bibinfo {volume} {B654}},\
  \bibinfo {pages} {110} (\bibinfo {year} {2007})},\ \Eprint
  {http://arxiv.org/abs/0707.2463}{arXiv:0707.2463 [hep-ph]}\BibitemShut
  {NoStop}%
\bibitem [{\citenamefont {Hamaguchi}\ \emph {et~al.}(2009)\citenamefont
  {Hamaguchi}, \citenamefont {Nakamura}, \citenamefont {Shirai},\ and\
  \citenamefont {Yanagida}}]{Hamaguchi:2008rv}%
  \BibitemOpen
  \bibfield  {author} {\bibinfo {author} {\bibfnamefont {K.}~\bibnamefont
  {Hamaguchi}}, \bibinfo {author} {\bibfnamefont {E.}~\bibnamefont {Nakamura}},
  \bibinfo {author} {\bibfnamefont {S.}~\bibnamefont {Shirai}}, \ and\ \bibinfo
  {author} {\bibfnamefont {T.~T.}\ \bibnamefont {Yanagida}},\ }\href {\doibase
  10.1016/j.physletb.2009.03.025} {\bibfield  {journal} {\bibinfo  {journal}
  {Phys. Lett.}\ }\textbf {\bibinfo {volume} {B674}},\ \bibinfo {pages} {299}
  (\bibinfo {year} {2009})},\ \Eprint
  {http://arxiv.org/abs/0811.0737}{arXiv:0811.0737 [hep-ph]}\BibitemShut
  {NoStop}%
\bibitem [{\citenamefont {Hamaguchi}\ \emph {et~al.}(2010)\citenamefont
  {Hamaguchi}, \citenamefont {Nakamura}, \citenamefont {Shirai},\ and\
  \citenamefont {Yanagida}}]{Hamaguchi:2009db}%
  \BibitemOpen
  \bibfield  {author} {\bibinfo {author} {\bibfnamefont {K.}~\bibnamefont
  {Hamaguchi}}, \bibinfo {author} {\bibfnamefont {E.}~\bibnamefont {Nakamura}},
  \bibinfo {author} {\bibfnamefont {S.}~\bibnamefont {Shirai}}, \ and\ \bibinfo
  {author} {\bibfnamefont {T.~T.}\ \bibnamefont {Yanagida}},\ }\href {\doibase
  10.1007/JHEP04(2010)119} {\bibfield  {journal} {\bibinfo  {journal} {JHEP}\
  }\textbf {\bibinfo {volume} {04}},\ \bibinfo {pages} {119} (\bibinfo {year}
  {2010})},\ \Eprint {http://arxiv.org/abs/0912.1683}{arXiv:0912.1683
  [hep-ph]}\BibitemShut {NoStop}%
\bibitem [{\citenamefont {Murayama}\ and\ \citenamefont
  {Shu}(2010)}]{Murayama:2009nj}%
  \BibitemOpen
  \bibfield  {author} {\bibinfo {author} {\bibfnamefont {H.}~\bibnamefont
  {Murayama}}\ and\ \bibinfo {author} {\bibfnamefont {J.}~\bibnamefont {Shu}},\
  }\href {\doibase 10.1016/j.physletb.2010.02.037} {\bibfield  {journal}
  {\bibinfo  {journal} {Phys. Lett.}\ }\textbf {\bibinfo {volume} {B686}},\
  \bibinfo {pages} {162} (\bibinfo {year} {2010})},\ \Eprint
  {http://arxiv.org/abs/0905.1720}{arXiv:0905.1720 [hep-ph]}\BibitemShut
  {NoStop}%
\bibitem [{\citenamefont {Hambye}\ and\ \citenamefont
  {Tytgat}(2010)}]{Hambye:2009fg}%
  \BibitemOpen
  \bibfield  {author} {\bibinfo {author} {\bibfnamefont {T.}~\bibnamefont
  {Hambye}}\ and\ \bibinfo {author} {\bibfnamefont {M.~H.~G.}\ \bibnamefont
  {Tytgat}},\ }\href {\doibase 10.1016/j.physletb.2009.11.050} {\bibfield
  {journal} {\bibinfo  {journal} {Phys. Lett.}\ }\textbf {\bibinfo {volume}
  {B683}},\ \bibinfo {pages} {39} (\bibinfo {year} {2010})},\ \Eprint
  {http://arxiv.org/abs/0907.1007}{arXiv:0907.1007 [hep-ph]}\BibitemShut
  {NoStop}%
\bibitem [{\citenamefont {Antipin}\ \emph
  {et~al.}(2015{\natexlab{a}})\citenamefont {Antipin}, \citenamefont {Redi},\
  and\ \citenamefont {Strumia}}]{Antipin:2014qva}%
  \BibitemOpen
  \bibfield  {author} {\bibinfo {author} {\bibfnamefont {O.}~\bibnamefont
  {Antipin}}, \bibinfo {author} {\bibfnamefont {M.}~\bibnamefont {Redi}}, \
  and\ \bibinfo {author} {\bibfnamefont {A.}~\bibnamefont {Strumia}},\ }\href
  {\doibase 10.1007/JHEP01(2015)157} {\bibfield  {journal} {\bibinfo  {journal}
  {JHEP}\ }\textbf {\bibinfo {volume} {01}},\ \bibinfo {pages} {157} (\bibinfo
  {year} {2015}{\natexlab{a}})},\ \Eprint
  {http://arxiv.org/abs/1410.1817}{arXiv:1410.1817 [hep-ph]}\BibitemShut
  {NoStop}%
\bibitem [{\citenamefont {Antipin}\ \emph
  {et~al.}(2015{\natexlab{b}})\citenamefont {Antipin}, \citenamefont {Redi},
  \citenamefont {Strumia},\ and\ \citenamefont {Vigiani}}]{Antipin:2015xia}%
  \BibitemOpen
  \bibfield  {author} {\bibinfo {author} {\bibfnamefont {O.}~\bibnamefont
  {Antipin}}, \bibinfo {author} {\bibfnamefont {M.}~\bibnamefont {Redi}},
  \bibinfo {author} {\bibfnamefont {A.}~\bibnamefont {Strumia}}, \ and\
  \bibinfo {author} {\bibfnamefont {E.}~\bibnamefont {Vigiani}},\ }\href
  {\doibase 10.1007/JHEP07(2015)039} {\bibfield  {journal} {\bibinfo  {journal}
  {JHEP}\ }\textbf {\bibinfo {volume} {07}},\ \bibinfo {pages} {039} (\bibinfo
  {year} {2015}{\natexlab{b}})},\ \Eprint
  {http://arxiv.org/abs/1503.08749}{arXiv:1503.08749 [hep-ph]}\BibitemShut
  {NoStop}%
\bibitem [{\citenamefont {Gross}\ \emph {et~al.}(2019)\citenamefont {Gross},
  \citenamefont {Mitridate}, \citenamefont {Redi}, \citenamefont {Smirnov},\
  and\ \citenamefont {Strumia}}]{Gross:2018zha}%
  \BibitemOpen
  \bibfield  {author} {\bibinfo {author} {\bibfnamefont {C.}~\bibnamefont
  {Gross}}, \bibinfo {author} {\bibfnamefont {A.}~\bibnamefont {Mitridate}},
  \bibinfo {author} {\bibfnamefont {M.}~\bibnamefont {Redi}}, \bibinfo {author}
  {\bibfnamefont {J.}~\bibnamefont {Smirnov}}, \ and\ \bibinfo {author}
  {\bibfnamefont {A.}~\bibnamefont {Strumia}},\ }\href {\doibase
  10.1103/PhysRevD.99.016024} {\bibfield  {journal} {\bibinfo  {journal} {Phys.
  Rev.}\ }\textbf {\bibinfo {volume} {D99}},\ \bibinfo {pages} {016024}
  (\bibinfo {year} {2019})},\ \Eprint
  {http://arxiv.org/abs/1811.08418}{arXiv:1811.08418 [hep-ph]}\BibitemShut
  {NoStop}%
\bibitem [{\citenamefont {Fukuda}\ \emph {et~al.}(2019)\citenamefont {Fukuda},
  \citenamefont {Luo},\ and\ \citenamefont {Shirai}}]{Fukuda:2018ufg}%
  \BibitemOpen
  \bibfield  {author} {\bibinfo {author} {\bibfnamefont {H.}~\bibnamefont
  {Fukuda}}, \bibinfo {author} {\bibfnamefont {F.}~\bibnamefont {Luo}}, \ and\
  \bibinfo {author} {\bibfnamefont {S.}~\bibnamefont {Shirai}},\ }\href
  {\doibase 10.1007/JHEP04(2019)107} {\bibfield  {journal} {\bibinfo  {journal}
  {JHEP}\ }\textbf {\bibinfo {volume} {04}},\ \bibinfo {pages} {107} (\bibinfo
  {year} {2019})},\ \Eprint {http://arxiv.org/abs/1812.02066}{arXiv:1812.02066
  [hep-ph]}\BibitemShut {NoStop}%
\bibitem [{\citenamefont {Bernstein}\ \emph {et~al.}(1985)\citenamefont
  {Bernstein}, \citenamefont {Brown},\ and\ \citenamefont
  {Feinberg}}]{Bernstein:1985th}%
  \BibitemOpen
  \bibfield  {author} {\bibinfo {author} {\bibfnamefont {J.}~\bibnamefont
  {Bernstein}}, \bibinfo {author} {\bibfnamefont {L.~S.}\ \bibnamefont
  {Brown}}, \ and\ \bibinfo {author} {\bibfnamefont {G.}~\bibnamefont
  {Feinberg}},\ }\href {\doibase 10.1103/PhysRevD.32.3261} {\bibfield
  {journal} {\bibinfo  {journal} {Phys. Rev.}\ }\textbf {\bibinfo {volume}
  {D32}},\ \bibinfo {pages} {3261} (\bibinfo {year} {1985})}\BibitemShut
  {NoStop}%
\bibitem [{\citenamefont {Srednicki}\ \emph {et~al.}(1988)\citenamefont
  {Srednicki}, \citenamefont {Watkins},\ and\ \citenamefont
  {Olive}}]{Srednicki:1988ce}%
  \BibitemOpen
  \bibfield  {author} {\bibinfo {author} {\bibfnamefont {M.}~\bibnamefont
  {Srednicki}}, \bibinfo {author} {\bibfnamefont {R.}~\bibnamefont {Watkins}},
  \ and\ \bibinfo {author} {\bibfnamefont {K.~A.}\ \bibnamefont {Olive}},\
  }\href {\doibase 10.1016/0550-3213(88)90099-5} {\bibfield  {journal}
  {\bibinfo  {journal} {Nucl. Phys.}\ }\textbf {\bibinfo {volume} {B310}},\
  \bibinfo {pages} {693} (\bibinfo {year} {1988})},\ \bibinfo {note}
  {[,247(1988)]}\BibitemShut {NoStop}%
\bibitem [{\citenamefont {Cirelli}\ \emph {et~al.}(2006)\citenamefont
  {Cirelli}, \citenamefont {Fornengo},\ and\ \citenamefont
  {Strumia}}]{Cirelli:2005uq}%
  \BibitemOpen
  \bibfield  {author} {\bibinfo {author} {\bibfnamefont {M.}~\bibnamefont
  {Cirelli}}, \bibinfo {author} {\bibfnamefont {N.}~\bibnamefont {Fornengo}}, \
  and\ \bibinfo {author} {\bibfnamefont {A.}~\bibnamefont {Strumia}},\ }\href
  {\doibase 10.1016/j.nuclphysb.2006.07.012} {\bibfield  {journal} {\bibinfo
  {journal} {Nucl. Phys.}\ }\textbf {\bibinfo {volume} {B753}},\ \bibinfo
  {pages} {178} (\bibinfo {year} {2006})},\ \Eprint
  {http://arxiv.org/abs/hep-ph/0512090}{arXiv:hep-ph/0512090
  [hep-ph]}\BibitemShut {NoStop}%
\bibitem [{\citenamefont {Cirelli}\ \emph {et~al.}(2007)\citenamefont
  {Cirelli}, \citenamefont {Strumia},\ and\ \citenamefont
  {Tamburini}}]{Cirelli:2007xd}%
  \BibitemOpen
  \bibfield  {author} {\bibinfo {author} {\bibfnamefont {M.}~\bibnamefont
  {Cirelli}}, \bibinfo {author} {\bibfnamefont {A.}~\bibnamefont {Strumia}}, \
  and\ \bibinfo {author} {\bibfnamefont {M.}~\bibnamefont {Tamburini}},\ }\href
  {\doibase 10.1016/j.nuclphysb.2007.07.023} {\bibfield  {journal} {\bibinfo
  {journal} {Nucl. Phys.}\ }\textbf {\bibinfo {volume} {B787}},\ \bibinfo
  {pages} {152} (\bibinfo {year} {2007})},\ \Eprint
  {http://arxiv.org/abs/0706.4071}{arXiv:0706.4071 [hep-ph]}\BibitemShut
  {NoStop}%
\bibitem [{\citenamefont {Cirelli}\ and\ \citenamefont
  {Strumia}(2009)}]{Cirelli:2009uv}%
  \BibitemOpen
  \bibfield  {author} {\bibinfo {author} {\bibfnamefont {M.}~\bibnamefont
  {Cirelli}}\ and\ \bibinfo {author} {\bibfnamefont {A.}~\bibnamefont
  {Strumia}},\ }\href {\doibase 10.1088/1367-2630/11/10/105005} {\bibfield
  {journal} {\bibinfo  {journal} {New J. Phys.}\ }\textbf {\bibinfo {volume}
  {11}},\ \bibinfo {pages} {105005} (\bibinfo {year} {2009})},\ \Eprint
  {http://arxiv.org/abs/0903.3381}{arXiv:0903.3381 [hep-ph]}\BibitemShut
  {NoStop}%
\bibitem [{\citenamefont {Nagata}\ and\ \citenamefont
  {Shirai}(2015)}]{Nagata:2014aoa}%
  \BibitemOpen
  \bibfield  {author} {\bibinfo {author} {\bibfnamefont {N.}~\bibnamefont
  {Nagata}}\ and\ \bibinfo {author} {\bibfnamefont {S.}~\bibnamefont
  {Shirai}},\ }\href {\doibase 10.1103/PhysRevD.91.055035} {\bibfield
  {journal} {\bibinfo  {journal} {Phys. Rev.}\ }\textbf {\bibinfo {volume}
  {D91}},\ \bibinfo {pages} {055035} (\bibinfo {year} {2015})},\ \Eprint
  {http://arxiv.org/abs/1411.0752}{arXiv:1411.0752 [hep-ph]}\BibitemShut
  {NoStop}%
\bibitem [{\citenamefont {Berggren}\ \emph {et~al.}(2013)\citenamefont
  {Berggren}, \citenamefont {Brummer}, \citenamefont {List}, \citenamefont
  {Moortgat-Pick}, \citenamefont {Robens}, \citenamefont {Rolbiecki},\ and\
  \citenamefont {Sert}}]{Berggren:2013vfa}%
  \BibitemOpen
  \bibfield  {author} {\bibinfo {author} {\bibfnamefont {M.}~\bibnamefont
  {Berggren}}, \bibinfo {author} {\bibfnamefont {F.}~\bibnamefont {Brummer}},
  \bibinfo {author} {\bibfnamefont {J.}~\bibnamefont {List}}, \bibinfo {author}
  {\bibfnamefont {G.}~\bibnamefont {Moortgat-Pick}}, \bibinfo {author}
  {\bibfnamefont {T.}~\bibnamefont {Robens}}, \bibinfo {author} {\bibfnamefont
  {K.}~\bibnamefont {Rolbiecki}}, \ and\ \bibinfo {author} {\bibfnamefont
  {H.}~\bibnamefont {Sert}},\ }\href {\doibase 10.1140/epjc/s10052-013-2660-y}
  {\bibfield  {journal} {\bibinfo  {journal} {Eur. Phys. J.}\ }\textbf
  {\bibinfo {volume} {C73}},\ \bibinfo {pages} {2660} (\bibinfo {year}
  {2013})},\ \Eprint {http://arxiv.org/abs/1307.3566}{arXiv:1307.3566
  [hep-ph]}\BibitemShut {NoStop}%
\bibitem [{\citenamefont {Baer}\ \emph {et~al.}(2014)\citenamefont {Baer},
  \citenamefont {Barger}, \citenamefont {Mickelson}, \citenamefont
  {Mustafayev},\ and\ \citenamefont {Tata}}]{Baer:2014yta}%
  \BibitemOpen
  \bibfield  {author} {\bibinfo {author} {\bibfnamefont {H.}~\bibnamefont
  {Baer}}, \bibinfo {author} {\bibfnamefont {V.}~\bibnamefont {Barger}},
  \bibinfo {author} {\bibfnamefont {D.}~\bibnamefont {Mickelson}}, \bibinfo
  {author} {\bibfnamefont {A.}~\bibnamefont {Mustafayev}}, \ and\ \bibinfo
  {author} {\bibfnamefont {X.}~\bibnamefont {Tata}},\ }\href {\doibase
  10.1007/JHEP06(2014)172} {\bibfield  {journal} {\bibinfo  {journal} {JHEP}\
  }\textbf {\bibinfo {volume} {06}},\ \bibinfo {pages} {172} (\bibinfo {year}
  {2014})},\ \Eprint {http://arxiv.org/abs/1404.7510}{arXiv:1404.7510
  [hep-ph]}\BibitemShut {NoStop}%
\bibitem [{\citenamefont {Fujii}\ \emph {et~al.}(2017)\citenamefont {Fujii}
  \emph {et~al.}}]{Fujii:2017ekh}%
  \BibitemOpen
  \bibfield  {author} {\bibinfo {author} {\bibfnamefont {K.}~\bibnamefont
  {Fujii}} \emph {et~al.},\ }\href@noop {} {\  (\bibinfo {year} {2017})},\
  \Eprint {http://arxiv.org/abs/1702.05333}{arXiv:1702.05333
  [hep-ph]}\BibitemShut {NoStop}%
\bibitem [{\citenamefont {Polesello}\ and\ \citenamefont
  {Tovey}(2004)}]{Polesello:2004qy}%
  \BibitemOpen
  \bibfield  {author} {\bibinfo {author} {\bibfnamefont {G.}~\bibnamefont
  {Polesello}}\ and\ \bibinfo {author} {\bibfnamefont {D.~R.}\ \bibnamefont
  {Tovey}},\ }\href {\doibase 10.1088/1126-6708/2004/05/071} {\bibfield
  {journal} {\bibinfo  {journal} {JHEP}\ }\textbf {\bibinfo {volume} {05}},\
  \bibinfo {pages} {071} (\bibinfo {year} {2004})},\ \Eprint
  {http://arxiv.org/abs/hep-ph/0403047}{arXiv:hep-ph/0403047
  [hep-ph]}\BibitemShut {NoStop}%
\bibitem [{\citenamefont {Battaglia}\ \emph {et~al.}(2004)\citenamefont
  {Battaglia}, \citenamefont {Hinchliffe},\ and\ \citenamefont
  {Tovey}}]{Battaglia:2004mp}%
  \BibitemOpen
  \bibfield  {author} {\bibinfo {author} {\bibfnamefont {M.}~\bibnamefont
  {Battaglia}}, \bibinfo {author} {\bibfnamefont {I.}~\bibnamefont
  {Hinchliffe}}, \ and\ \bibinfo {author} {\bibfnamefont {D.}~\bibnamefont
  {Tovey}},\ }\href {\doibase 10.1088/0954-3899/30/10/R01} {\bibfield
  {journal} {\bibinfo  {journal} {J. Phys.}\ }\textbf {\bibinfo {volume}
  {G30}},\ \bibinfo {pages} {R217} (\bibinfo {year} {2004})},\ \Eprint
  {http://arxiv.org/abs/hep-ph/0406147}{arXiv:hep-ph/0406147
  [hep-ph]}\BibitemShut {NoStop}%
\bibitem [{\citenamefont {Weiglein}\ \emph {et~al.}(2006)\citenamefont
  {Weiglein} \emph {et~al.}}]{Weiglein:2004hn}%
  \BibitemOpen
  \bibfield  {author} {\bibinfo {author} {\bibfnamefont {G.}~\bibnamefont
  {Weiglein}} \emph {et~al.} (\bibinfo {collaboration} {LHC/LC Study Group}),\
  }\href {\doibase 10.1016/j.physrep.2005.12.003} {\bibfield  {journal}
  {\bibinfo  {journal} {Phys. Rept.}\ }\textbf {\bibinfo {volume} {426}},\
  \bibinfo {pages} {47} (\bibinfo {year} {2006})},\ \Eprint
  {http://arxiv.org/abs/hep-ph/0410364}{arXiv:hep-ph/0410364
  [hep-ph]}\BibitemShut {NoStop}%
\bibitem [{\citenamefont {Baltz}\ \emph {et~al.}(2006)\citenamefont {Baltz},
  \citenamefont {Battaglia}, \citenamefont {Peskin},\ and\ \citenamefont
  {Wizansky}}]{Baltz:2006fm}%
  \BibitemOpen
  \bibfield  {author} {\bibinfo {author} {\bibfnamefont {E.~A.}\ \bibnamefont
  {Baltz}}, \bibinfo {author} {\bibfnamefont {M.}~\bibnamefont {Battaglia}},
  \bibinfo {author} {\bibfnamefont {M.~E.}\ \bibnamefont {Peskin}}, \ and\
  \bibinfo {author} {\bibfnamefont {T.}~\bibnamefont {Wizansky}},\ }\href
  {\doibase 10.1103/PhysRevD.74.103521} {\bibfield  {journal} {\bibinfo
  {journal} {Phys. Rev.}\ }\textbf {\bibinfo {volume} {D74}},\ \bibinfo {pages}
  {103521} (\bibinfo {year} {2006})},\ \Eprint
  {http://arxiv.org/abs/hep-ph/0602187}{arXiv:hep-ph/0602187
  [hep-ph]}\BibitemShut {NoStop}%
\bibitem [{\citenamefont {Asakawa}\ \emph {et~al.}(2009)\citenamefont
  {Asakawa}, \citenamefont {Asano}, \citenamefont {Fujii}, \citenamefont
  {Kusano}, \citenamefont {Matsumoto}, \citenamefont {Sasaki}, \citenamefont
  {Takubo},\ and\ \citenamefont {Yamamoto}}]{Asakawa:2009qb}%
  \BibitemOpen
  \bibfield  {author} {\bibinfo {author} {\bibfnamefont {E.}~\bibnamefont
  {Asakawa}}, \bibinfo {author} {\bibfnamefont {M.}~\bibnamefont {Asano}},
  \bibinfo {author} {\bibfnamefont {K.}~\bibnamefont {Fujii}}, \bibinfo
  {author} {\bibfnamefont {T.}~\bibnamefont {Kusano}}, \bibinfo {author}
  {\bibfnamefont {S.}~\bibnamefont {Matsumoto}}, \bibinfo {author}
  {\bibfnamefont {R.}~\bibnamefont {Sasaki}}, \bibinfo {author} {\bibfnamefont
  {Y.}~\bibnamefont {Takubo}}, \ and\ \bibinfo {author} {\bibfnamefont
  {H.}~\bibnamefont {Yamamoto}},\ }\href {\doibase 10.1103/PhysRevD.79.075013}
  {\bibfield  {journal} {\bibinfo  {journal} {Phys. Rev.}\ }\textbf {\bibinfo
  {volume} {D79}},\ \bibinfo {pages} {075013} (\bibinfo {year} {2009})},\
  \Eprint {http://arxiv.org/abs/0901.1081}{arXiv:0901.1081
  [hep-ph]}\BibitemShut {NoStop}%
\bibitem [{\citenamefont {Hisano}\ \emph {et~al.}(2004)\citenamefont {Hisano},
  \citenamefont {Matsumoto},\ and\ \citenamefont {Nojiri}}]{Hisano:2003ec}%
  \BibitemOpen
  \bibfield  {author} {\bibinfo {author} {\bibfnamefont {J.}~\bibnamefont
  {Hisano}}, \bibinfo {author} {\bibfnamefont {S.}~\bibnamefont {Matsumoto}}, \
  and\ \bibinfo {author} {\bibfnamefont {M.~M.}\ \bibnamefont {Nojiri}},\
  }\href {\doibase 10.1103/PhysRevLett.92.031303} {\bibfield  {journal}
  {\bibinfo  {journal} {Phys. Rev. Lett.}\ }\textbf {\bibinfo {volume} {92}},\
  \bibinfo {pages} {031303} (\bibinfo {year} {2004})},\ \Eprint
  {http://arxiv.org/abs/hep-ph/0307216}{arXiv:hep-ph/0307216
  [hep-ph]}\BibitemShut {NoStop}%
\bibitem [{\citenamefont {Hisano}\ \emph {et~al.}(2007)\citenamefont {Hisano},
  \citenamefont {Matsumoto}, \citenamefont {Nagai}, \citenamefont {Saito},\
  and\ \citenamefont {Senami}}]{Hisano:2006nn}%
  \BibitemOpen
  \bibfield  {author} {\bibinfo {author} {\bibfnamefont {J.}~\bibnamefont
  {Hisano}}, \bibinfo {author} {\bibfnamefont {S.}~\bibnamefont {Matsumoto}},
  \bibinfo {author} {\bibfnamefont {M.}~\bibnamefont {Nagai}}, \bibinfo
  {author} {\bibfnamefont {O.}~\bibnamefont {Saito}}, \ and\ \bibinfo {author}
  {\bibfnamefont {M.}~\bibnamefont {Senami}},\ }\href {\doibase
  10.1016/j.physletb.2007.01.012} {\bibfield  {journal} {\bibinfo  {journal}
  {Phys. Lett.}\ }\textbf {\bibinfo {volume} {B646}},\ \bibinfo {pages} {34}
  (\bibinfo {year} {2007})},\ \Eprint
  {http://arxiv.org/abs/hep-ph/0610249}{arXiv:hep-ph/0610249
  [hep-ph]}\BibitemShut {NoStop}%
\bibitem [{\citenamefont {Pospelov}\ and\ \citenamefont
  {Ritz}(2009)}]{Pospelov:2008jd}%
  \BibitemOpen
  \bibfield  {author} {\bibinfo {author} {\bibfnamefont {M.}~\bibnamefont
  {Pospelov}}\ and\ \bibinfo {author} {\bibfnamefont {A.}~\bibnamefont
  {Ritz}},\ }\href {\doibase 10.1016/j.physletb.2008.12.012} {\bibfield
  {journal} {\bibinfo  {journal} {Phys. Lett.}\ }\textbf {\bibinfo {volume}
  {B671}},\ \bibinfo {pages} {391} (\bibinfo {year} {2009})},\ \Eprint
  {http://arxiv.org/abs/0810.1502}{arXiv:0810.1502 [hep-ph]}\BibitemShut
  {NoStop}%
\bibitem [{\citenamefont {March-Russell}\ and\ \citenamefont
  {West}(2009)}]{MarchRussell:2008tu}%
  \BibitemOpen
  \bibfield  {author} {\bibinfo {author} {\bibfnamefont {J.~D.}\ \bibnamefont
  {March-Russell}}\ and\ \bibinfo {author} {\bibfnamefont {S.~M.}\ \bibnamefont
  {West}},\ }\href {\doibase 10.1016/j.physletb.2009.04.010} {\bibfield
  {journal} {\bibinfo  {journal} {Phys. Lett.}\ }\textbf {\bibinfo {volume}
  {B676}},\ \bibinfo {pages} {133} (\bibinfo {year} {2009})},\ \Eprint
  {http://arxiv.org/abs/0812.0559}{arXiv:0812.0559 [astro-ph]}\BibitemShut
  {NoStop}%
\bibitem [{\citenamefont {Shepherd}\ \emph {et~al.}(2009)\citenamefont
  {Shepherd}, \citenamefont {Tait},\ and\ \citenamefont
  {Zaharijas}}]{Shepherd:2009sa}%
  \BibitemOpen
  \bibfield  {author} {\bibinfo {author} {\bibfnamefont {W.}~\bibnamefont
  {Shepherd}}, \bibinfo {author} {\bibfnamefont {T.~M.~P.}\ \bibnamefont
  {Tait}}, \ and\ \bibinfo {author} {\bibfnamefont {G.}~\bibnamefont
  {Zaharijas}},\ }\href {\doibase 10.1103/PhysRevD.79.055022} {\bibfield
  {journal} {\bibinfo  {journal} {Phys. Rev.}\ }\textbf {\bibinfo {volume}
  {D79}},\ \bibinfo {pages} {055022} (\bibinfo {year} {2009})},\ \Eprint
  {http://arxiv.org/abs/0901.2125}{arXiv:0901.2125 [hep-ph]}\BibitemShut
  {NoStop}%
\bibitem [{\citenamefont {Feng}\ \emph {et~al.}(2009)\citenamefont {Feng},
  \citenamefont {Kaplinghat}, \citenamefont {Tu},\ and\ \citenamefont
  {Yu}}]{Feng:2009mn}%
  \BibitemOpen
  \bibfield  {author} {\bibinfo {author} {\bibfnamefont {J.~L.}\ \bibnamefont
  {Feng}}, \bibinfo {author} {\bibfnamefont {M.}~\bibnamefont {Kaplinghat}},
  \bibinfo {author} {\bibfnamefont {H.}~\bibnamefont {Tu}}, \ and\ \bibinfo
  {author} {\bibfnamefont {H.-B.}\ \bibnamefont {Yu}},\ }\href {\doibase
  10.1088/1475-7516/2009/07/004} {\bibfield  {journal} {\bibinfo  {journal}
  {JCAP}\ }\textbf {\bibinfo {volume} {07}},\ \bibinfo {pages} {004} (\bibinfo
  {year} {2009})},\ \Eprint {http://arxiv.org/abs/0905.3039}{arXiv:0905.3039
  [hep-ph]}\BibitemShut {NoStop}%
\bibitem [{\citenamefont {von Harling}\ and\ \citenamefont
  {Petraki}(2014)}]{vonHarling:2014kha}%
  \BibitemOpen
  \bibfield  {author} {\bibinfo {author} {\bibfnamefont {B.}~\bibnamefont {von
  Harling}}\ and\ \bibinfo {author} {\bibfnamefont {K.}~\bibnamefont
  {Petraki}},\ }\href {\doibase 10.1088/1475-7516/2014/12/033} {\bibfield
  {journal} {\bibinfo  {journal} {JCAP}\ }\textbf {\bibinfo {volume} {12}},\
  \bibinfo {pages} {033} (\bibinfo {year} {2014})},\ \Eprint
  {http://arxiv.org/abs/1407.7874}{arXiv:1407.7874 [hep-ph]}\BibitemShut
  {NoStop}%
\bibitem [{\citenamefont {Berger}\ \emph {et~al.}(2008)\citenamefont {Berger},
  \citenamefont {Covi}, \citenamefont {Kraml},\ and\ \citenamefont
  {Palorini}}]{Berger:2008ti}%
  \BibitemOpen
  \bibfield  {author} {\bibinfo {author} {\bibfnamefont {C.~F.}\ \bibnamefont
  {Berger}}, \bibinfo {author} {\bibfnamefont {L.}~\bibnamefont {Covi}},
  \bibinfo {author} {\bibfnamefont {S.}~\bibnamefont {Kraml}}, \ and\ \bibinfo
  {author} {\bibfnamefont {F.}~\bibnamefont {Palorini}},\ }\href {\doibase
  10.1088/1475-7516/2008/10/005} {\bibfield  {journal} {\bibinfo  {journal}
  {JCAP}\ }\textbf {\bibinfo {volume} {0810}},\ \bibinfo {pages} {005}
  (\bibinfo {year} {2008})},\ \Eprint
  {http://arxiv.org/abs/0807.0211}{arXiv:0807.0211 [hep-ph]}\BibitemShut
  {NoStop}%
\bibitem [{\citenamefont {Blum}\ \emph {et~al.}(2016)\citenamefont {Blum},
  \citenamefont {Sato},\ and\ \citenamefont {Slatyer}}]{Blum:2016nrz}%
  \BibitemOpen
  \bibfield  {author} {\bibinfo {author} {\bibfnamefont {K.}~\bibnamefont
  {Blum}}, \bibinfo {author} {\bibfnamefont {R.}~\bibnamefont {Sato}}, \ and\
  \bibinfo {author} {\bibfnamefont {T.~R.}\ \bibnamefont {Slatyer}},\ }\href
  {\doibase 10.1088/1475-7516/2016/06/021} {\bibfield  {journal} {\bibinfo
  {journal} {JCAP}\ }\textbf {\bibinfo {volume} {1606}},\ \bibinfo {pages}
  {021} (\bibinfo {year} {2016})},\ \Eprint
  {http://arxiv.org/abs/1603.01383}{arXiv:1603.01383 [hep-ph]}\BibitemShut
  {NoStop}%
\bibitem [{\citenamefont {Mitridate}\ \emph {et~al.}(2017)\citenamefont
  {Mitridate}, \citenamefont {Redi}, \citenamefont {Smirnov},\ and\
  \citenamefont {Strumia}}]{Mitridate:2017izz}%
  \BibitemOpen
  \bibfield  {author} {\bibinfo {author} {\bibfnamefont {A.}~\bibnamefont
  {Mitridate}}, \bibinfo {author} {\bibfnamefont {M.}~\bibnamefont {Redi}},
  \bibinfo {author} {\bibfnamefont {J.}~\bibnamefont {Smirnov}}, \ and\
  \bibinfo {author} {\bibfnamefont {A.}~\bibnamefont {Strumia}},\ }\href
  {\doibase 10.1088/1475-7516/2017/05/006} {\bibfield  {journal} {\bibinfo
  {journal} {JCAP}\ }\textbf {\bibinfo {volume} {1705}},\ \bibinfo {pages}
  {006} (\bibinfo {year} {2017})},\ \Eprint
  {http://arxiv.org/abs/1702.01141}{arXiv:1702.01141 [hep-ph]}\BibitemShut
  {NoStop}%
\bibitem [{\citenamefont {Baldes}\ and\ \citenamefont
  {Petraki}(2017)}]{Baldes:2017gzw}%
  \BibitemOpen
  \bibfield  {author} {\bibinfo {author} {\bibfnamefont {I.}~\bibnamefont
  {Baldes}}\ and\ \bibinfo {author} {\bibfnamefont {K.}~\bibnamefont
  {Petraki}},\ }\href {\doibase 10.1088/1475-7516/2017/09/028} {\bibfield
  {journal} {\bibinfo  {journal} {JCAP}\ }\textbf {\bibinfo {volume} {09}},\
  \bibinfo {pages} {028} (\bibinfo {year} {2017})},\ \Eprint
  {http://arxiv.org/abs/1703.00478}{arXiv:1703.00478 [hep-ph]}\BibitemShut
  {NoStop}%
\bibitem [{\citenamefont {Harz}\ and\ \citenamefont
  {Petraki}(2018{\natexlab{a}})}]{Harz:2017dlj}%
  \BibitemOpen
  \bibfield  {author} {\bibinfo {author} {\bibfnamefont {J.}~\bibnamefont
  {Harz}}\ and\ \bibinfo {author} {\bibfnamefont {K.}~\bibnamefont {Petraki}},\
  }\href {\doibase 10.1103/PhysRevD.97.075041} {\bibfield  {journal} {\bibinfo
  {journal} {Phys. Rev. D}\ }\textbf {\bibinfo {volume} {97}},\ \bibinfo
  {pages} {075041} (\bibinfo {year} {2018}{\natexlab{a}})},\ \Eprint
  {http://arxiv.org/abs/1711.03552}{arXiv:1711.03552 [hep-ph]}\BibitemShut
  {NoStop}%
\bibitem [{\citenamefont {Harz}\ and\ \citenamefont
  {Petraki}(2018{\natexlab{b}})}]{Harz:2018csl}%
  \BibitemOpen
  \bibfield  {author} {\bibinfo {author} {\bibfnamefont {J.}~\bibnamefont
  {Harz}}\ and\ \bibinfo {author} {\bibfnamefont {K.}~\bibnamefont {Petraki}},\
  }\href {\doibase 10.1007/JHEP07(2018)096} {\bibfield  {journal} {\bibinfo
  {journal} {JHEP}\ }\textbf {\bibinfo {volume} {07}},\ \bibinfo {pages} {096}
  (\bibinfo {year} {2018}{\natexlab{b}})},\ \Eprint
  {http://arxiv.org/abs/1805.01200}{arXiv:1805.01200 [hep-ph]}\BibitemShut
  {NoStop}%
\bibitem [{\citenamefont {Harz}\ and\ \citenamefont
  {Petraki}(2019)}]{Harz:2019rro}%
  \BibitemOpen
  \bibfield  {author} {\bibinfo {author} {\bibfnamefont {J.}~\bibnamefont
  {Harz}}\ and\ \bibinfo {author} {\bibfnamefont {K.}~\bibnamefont {Petraki}},\
  }\href {\doibase 10.1007/JHEP04(2019)130} {\bibfield  {journal} {\bibinfo
  {journal} {JHEP}\ }\textbf {\bibinfo {volume} {04}},\ \bibinfo {pages} {130}
  (\bibinfo {year} {2019})},\ \Eprint
  {http://arxiv.org/abs/1901.10030}{arXiv:1901.10030 [hep-ph]}\BibitemShut
  {NoStop}%
\bibitem [{\citenamefont {Oncala}\ and\ \citenamefont
  {Petraki}(2020)}]{Oncala:2019yvj}%
  \BibitemOpen
  \bibfield  {author} {\bibinfo {author} {\bibfnamefont {R.}~\bibnamefont
  {Oncala}}\ and\ \bibinfo {author} {\bibfnamefont {K.}~\bibnamefont
  {Petraki}},\ }\href {\doibase 10.1007/JHEP02(2020)036} {\bibfield  {journal}
  {\bibinfo  {journal} {JHEP}\ }\textbf {\bibinfo {volume} {02}},\ \bibinfo
  {pages} {036} (\bibinfo {year} {2020})},\ \Eprint
  {http://arxiv.org/abs/1911.02605}{arXiv:1911.02605 [hep-ph]}\BibitemShut
  {NoStop}%
\bibitem [{\citenamefont {Biondini}\ and\ \citenamefont
  {Laine}(2017)}]{Biondini:2017ufr}%
  \BibitemOpen
  \bibfield  {author} {\bibinfo {author} {\bibfnamefont {S.}~\bibnamefont
  {Biondini}}\ and\ \bibinfo {author} {\bibfnamefont {M.}~\bibnamefont
  {Laine}},\ }\href {\doibase 10.1007/JHEP08(2017)047} {\bibfield  {journal}
  {\bibinfo  {journal} {JHEP}\ }\textbf {\bibinfo {volume} {08}},\ \bibinfo
  {pages} {047} (\bibinfo {year} {2017})},\ \Eprint
  {http://arxiv.org/abs/1706.01894}{arXiv:1706.01894 [hep-ph]}\BibitemShut
  {NoStop}%
\bibitem [{\citenamefont {Biondini}\ and\ \citenamefont
  {Laine}(2018)}]{Biondini:2018pwp}%
  \BibitemOpen
  \bibfield  {author} {\bibinfo {author} {\bibfnamefont {S.}~\bibnamefont
  {Biondini}}\ and\ \bibinfo {author} {\bibfnamefont {M.}~\bibnamefont
  {Laine}},\ }\href {\doibase 10.1007/JHEP04(2018)072} {\bibfield  {journal}
  {\bibinfo  {journal} {JHEP}\ }\textbf {\bibinfo {volume} {04}},\ \bibinfo
  {pages} {072} (\bibinfo {year} {2018})},\ \Eprint
  {http://arxiv.org/abs/1801.05821}{arXiv:1801.05821 [hep-ph]}\BibitemShut
  {NoStop}%
\bibitem [{\citenamefont {Biondini}(2018)}]{Biondini:2018xor}%
  \BibitemOpen
  \bibfield  {author} {\bibinfo {author} {\bibfnamefont {S.}~\bibnamefont
  {Biondini}},\ }\href {\doibase 10.1007/JHEP06(2018)104} {\bibfield  {journal}
  {\bibinfo  {journal} {JHEP}\ }\textbf {\bibinfo {volume} {06}},\ \bibinfo
  {pages} {104} (\bibinfo {year} {2018})},\ \Eprint
  {http://arxiv.org/abs/1805.00353}{arXiv:1805.00353 [hep-ph]}\BibitemShut
  {NoStop}%
\bibitem [{\citenamefont {Biondini}\ and\ \citenamefont
  {Vogl}(2019)}]{Biondini:2018ovz}%
  \BibitemOpen
  \bibfield  {author} {\bibinfo {author} {\bibfnamefont {S.}~\bibnamefont
  {Biondini}}\ and\ \bibinfo {author} {\bibfnamefont {S.}~\bibnamefont
  {Vogl}},\ }\href {\doibase 10.1007/JHEP02(2019)016} {\bibfield  {journal}
  {\bibinfo  {journal} {JHEP}\ }\textbf {\bibinfo {volume} {02}},\ \bibinfo
  {pages} {016} (\bibinfo {year} {2019})},\ \Eprint
  {http://arxiv.org/abs/1811.02581}{arXiv:1811.02581 [hep-ph]}\BibitemShut
  {NoStop}%
\bibitem [{\citenamefont {Binder}\ \emph {et~al.}(2018)\citenamefont {Binder},
  \citenamefont {Covi},\ and\ \citenamefont {Mukaida}}]{Binder:2018znk}%
  \BibitemOpen
  \bibfield  {author} {\bibinfo {author} {\bibfnamefont {T.}~\bibnamefont
  {Binder}}, \bibinfo {author} {\bibfnamefont {L.}~\bibnamefont {Covi}}, \ and\
  \bibinfo {author} {\bibfnamefont {K.}~\bibnamefont {Mukaida}},\ }\href
  {\doibase 10.1103/PhysRevD.98.115023} {\bibfield  {journal} {\bibinfo
  {journal} {Phys. Rev.}\ }\textbf {\bibinfo {volume} {D98}},\ \bibinfo {pages}
  {115023} (\bibinfo {year} {2018})},\ \Eprint
  {http://arxiv.org/abs/1808.06472}{arXiv:1808.06472 [hep-ph]}\BibitemShut
  {NoStop}%
\bibitem [{\citenamefont {Binder}\ \emph {et~al.}(2019)\citenamefont {Binder},
  \citenamefont {Mukaida},\ and\ \citenamefont {Petraki}}]{Binder:2019erp}%
  \BibitemOpen
  \bibfield  {author} {\bibinfo {author} {\bibfnamefont {T.}~\bibnamefont
  {Binder}}, \bibinfo {author} {\bibfnamefont {K.}~\bibnamefont {Mukaida}}, \
  and\ \bibinfo {author} {\bibfnamefont {K.}~\bibnamefont {Petraki}},\
  }\href@noop {} {\  (\bibinfo {year} {2019})},\ \Eprint
  {http://arxiv.org/abs/1910.11288}{arXiv:1910.11288 [hep-ph]}\BibitemShut
  {NoStop}%
\bibitem [{\citenamefont {Saikawa}\ and\ \citenamefont
  {Shirai}(2018)}]{Saikawa:2018rcs}%
  \BibitemOpen
  \bibfield  {author} {\bibinfo {author} {\bibfnamefont {K.}~\bibnamefont
  {Saikawa}}\ and\ \bibinfo {author} {\bibfnamefont {S.}~\bibnamefont
  {Shirai}},\ }\href {\doibase 10.1088/1475-7516/2018/05/035} {\bibfield
  {journal} {\bibinfo  {journal} {JCAP}\ }\textbf {\bibinfo {volume} {1805}},\
  \bibinfo {pages} {035} (\bibinfo {year} {2018})},\ \Eprint
  {http://arxiv.org/abs/1803.01038}{arXiv:1803.01038 [hep-ph]}\BibitemShut
  {NoStop}%
\bibitem [{\citenamefont {Mangano}\ \emph {et~al.}(2002)\citenamefont
  {Mangano}, \citenamefont {Miele}, \citenamefont {Pastor},\ and\ \citenamefont
  {Peloso}}]{Mangano:2001iu}%
  \BibitemOpen
  \bibfield  {author} {\bibinfo {author} {\bibfnamefont {G.}~\bibnamefont
  {Mangano}}, \bibinfo {author} {\bibfnamefont {G.}~\bibnamefont {Miele}},
  \bibinfo {author} {\bibfnamefont {S.}~\bibnamefont {Pastor}}, \ and\ \bibinfo
  {author} {\bibfnamefont {M.}~\bibnamefont {Peloso}},\ }\href {\doibase
  10.1016/S0370-2693(02)01622-2} {\bibfield  {journal} {\bibinfo  {journal}
  {Phys. Lett.}\ }\textbf {\bibinfo {volume} {B534}},\ \bibinfo {pages} {8}
  (\bibinfo {year} {2002})},\ \Eprint
  {http://arxiv.org/abs/astro-ph/0111408}{arXiv:astro-ph/0111408
  [astro-ph]}\BibitemShut {NoStop}%
\bibitem [{\citenamefont {Mangano}\ \emph {et~al.}(2005)\citenamefont
  {Mangano}, \citenamefont {Miele}, \citenamefont {Pastor}, \citenamefont
  {Pinto}, \citenamefont {Pisanti},\ and\ \citenamefont
  {Serpico}}]{Mangano:2005cc}%
  \BibitemOpen
  \bibfield  {author} {\bibinfo {author} {\bibfnamefont {G.}~\bibnamefont
  {Mangano}}, \bibinfo {author} {\bibfnamefont {G.}~\bibnamefont {Miele}},
  \bibinfo {author} {\bibfnamefont {S.}~\bibnamefont {Pastor}}, \bibinfo
  {author} {\bibfnamefont {T.}~\bibnamefont {Pinto}}, \bibinfo {author}
  {\bibfnamefont {O.}~\bibnamefont {Pisanti}}, \ and\ \bibinfo {author}
  {\bibfnamefont {P.~D.}\ \bibnamefont {Serpico}},\ }\href {\doibase
  10.1016/j.nuclphysb.2005.09.041} {\bibfield  {journal} {\bibinfo  {journal}
  {Nucl. Phys.}\ }\textbf {\bibinfo {volume} {B729}},\ \bibinfo {pages} {221}
  (\bibinfo {year} {2005})},\ \Eprint
  {http://arxiv.org/abs/hep-ph/0506164}{arXiv:hep-ph/0506164
  [hep-ph]}\BibitemShut {NoStop}%
\bibitem [{\citenamefont {de~Salas}\ and\ \citenamefont
  {Pastor}(2016)}]{deSalas:2016ztq}%
  \BibitemOpen
  \bibfield  {author} {\bibinfo {author} {\bibfnamefont {P.~F.}\ \bibnamefont
  {de~Salas}}\ and\ \bibinfo {author} {\bibfnamefont {S.}~\bibnamefont
  {Pastor}},\ }\href {\doibase 10.1088/1475-7516/2016/07/051} {\bibfield
  {journal} {\bibinfo  {journal} {JCAP}\ }\textbf {\bibinfo {volume} {1607}},\
  \bibinfo {pages} {051} (\bibinfo {year} {2016})},\ \Eprint
  {http://arxiv.org/abs/1606.06986}{arXiv:1606.06986 [hep-ph]}\BibitemShut
  {NoStop}%
\bibitem [{\citenamefont {Kolb}\ and\ \citenamefont
  {Turner}(1990)}]{Kolb:1990vq}%
  \BibitemOpen
  \bibfield  {author} {\bibinfo {author} {\bibfnamefont {E.~W.}\ \bibnamefont
  {Kolb}}\ and\ \bibinfo {author} {\bibfnamefont {M.~S.}\ \bibnamefont
  {Turner}},\ }\href@noop {} {\emph {\bibinfo {title} {{The Early
  Universe}}}},\ Front. Phys. ${\bf 69}$\ (\bibinfo {year} {1990})\BibitemShut
  {NoStop}%
\bibitem [{\citenamefont {Escudero~Abenza}(2020)}]{Escudero:2020dfa}%
  \BibitemOpen
  \bibfield  {author} {\bibinfo {author} {\bibfnamefont {M.}~\bibnamefont
  {Escudero~Abenza}},\ }\href@noop {} {\  (\bibinfo {year} {2020})},\ \Eprint
  {http://arxiv.org/abs/2001.04466}{arXiv:2001.04466 [hep-ph]}\BibitemShut
  {NoStop}%
\bibitem [{\citenamefont {Borsanyi}\ \emph {et~al.}(2010)\citenamefont
  {Borsanyi}, \citenamefont {Endrodi}, \citenamefont {Fodor}, \citenamefont
  {Jakovac}, \citenamefont {Katz}, \citenamefont {Krieg}, \citenamefont
  {Ratti},\ and\ \citenamefont {Szabo}}]{Borsanyi:2010cj}%
  \BibitemOpen
  \bibfield  {author} {\bibinfo {author} {\bibfnamefont {S.}~\bibnamefont
  {Borsanyi}}, \bibinfo {author} {\bibfnamefont {G.}~\bibnamefont {Endrodi}},
  \bibinfo {author} {\bibfnamefont {Z.}~\bibnamefont {Fodor}}, \bibinfo
  {author} {\bibfnamefont {A.}~\bibnamefont {Jakovac}}, \bibinfo {author}
  {\bibfnamefont {S.~D.}\ \bibnamefont {Katz}}, \bibinfo {author}
  {\bibfnamefont {S.}~\bibnamefont {Krieg}}, \bibinfo {author} {\bibfnamefont
  {C.}~\bibnamefont {Ratti}}, \ and\ \bibinfo {author} {\bibfnamefont {K.~K.}\
  \bibnamefont {Szabo}},\ }\href {\doibase 10.1007/JHEP11(2010)077} {\bibfield
  {journal} {\bibinfo  {journal} {JHEP}\ }\textbf {\bibinfo {volume} {11}},\
  \bibinfo {pages} {077} (\bibinfo {year} {2010})},\ \Eprint
  {http://arxiv.org/abs/1007.2580}{arXiv:1007.2580 [hep-lat]}\BibitemShut
  {NoStop}%
\bibitem [{\citenamefont {Borsanyi}\ \emph {et~al.}(2016)\citenamefont
  {Borsanyi} \emph {et~al.}}]{Borsanyi:2016ksw}%
  \BibitemOpen
  \bibfield  {author} {\bibinfo {author} {\bibfnamefont {S.}~\bibnamefont
  {Borsanyi}} \emph {et~al.},\ }\href {\doibase 10.1038/nature20115} {\bibfield
   {journal} {\bibinfo  {journal} {Nature}\ }\textbf {\bibinfo {volume}
  {539}},\ \bibinfo {pages} {69} (\bibinfo {year} {2016})},\ \Eprint
  {http://arxiv.org/abs/1606.07494}{arXiv:1606.07494 [hep-lat]}\BibitemShut
  {NoStop}%
\bibitem [{\citenamefont {Kajantie}\ \emph {et~al.}(2003)\citenamefont
  {Kajantie}, \citenamefont {Laine}, \citenamefont {Rummukainen},\ and\
  \citenamefont {Schroder}}]{Kajantie:2002wa}%
  \BibitemOpen
  \bibfield  {author} {\bibinfo {author} {\bibfnamefont {K.}~\bibnamefont
  {Kajantie}}, \bibinfo {author} {\bibfnamefont {M.}~\bibnamefont {Laine}},
  \bibinfo {author} {\bibfnamefont {K.}~\bibnamefont {Rummukainen}}, \ and\
  \bibinfo {author} {\bibfnamefont {Y.}~\bibnamefont {Schroder}},\ }\href
  {\doibase 10.1103/PhysRevD.67.105008} {\bibfield  {journal} {\bibinfo
  {journal} {Phys. Rev.}\ }\textbf {\bibinfo {volume} {D67}},\ \bibinfo {pages}
  {105008} (\bibinfo {year} {2003})},\ \Eprint
  {http://arxiv.org/abs/hep-ph/0211321}{arXiv:hep-ph/0211321
  [hep-ph]}\BibitemShut {NoStop}%
\bibitem [{\citenamefont {Laine}\ and\ \citenamefont
  {Schroder}(2006)}]{Laine:2006cp}%
  \BibitemOpen
  \bibfield  {author} {\bibinfo {author} {\bibfnamefont {M.}~\bibnamefont
  {Laine}}\ and\ \bibinfo {author} {\bibfnamefont {Y.}~\bibnamefont
  {Schroder}},\ }\href {\doibase 10.1103/PhysRevD.73.085009} {\bibfield
  {journal} {\bibinfo  {journal} {Phys. Rev.}\ }\textbf {\bibinfo {volume}
  {D73}},\ \bibinfo {pages} {085009} (\bibinfo {year} {2006})},\ \Eprint
  {http://arxiv.org/abs/hep-ph/0603048}{arXiv:hep-ph/0603048
  [hep-ph]}\BibitemShut {NoStop}%
\bibitem [{\citenamefont {Farakos}\ \emph {et~al.}(1994)\citenamefont
  {Farakos}, \citenamefont {Kajantie}, \citenamefont {Rummukainen},\ and\
  \citenamefont {Shaposhnikov}}]{Farakos:1994kx}%
  \BibitemOpen
  \bibfield  {author} {\bibinfo {author} {\bibfnamefont {K.}~\bibnamefont
  {Farakos}}, \bibinfo {author} {\bibfnamefont {K.}~\bibnamefont {Kajantie}},
  \bibinfo {author} {\bibfnamefont {K.}~\bibnamefont {Rummukainen}}, \ and\
  \bibinfo {author} {\bibfnamefont {M.~E.}\ \bibnamefont {Shaposhnikov}},\
  }\href {\doibase 10.1016/0550-3213(94)90173-2} {\bibfield  {journal}
  {\bibinfo  {journal} {Nucl. Phys.}\ }\textbf {\bibinfo {volume} {B425}},\
  \bibinfo {pages} {67} (\bibinfo {year} {1994})},\ \Eprint
  {http://arxiv.org/abs/hep-ph/9404201}{arXiv:hep-ph/9404201
  [hep-ph]}\BibitemShut {NoStop}%
\bibitem [{\citenamefont {Farakos}\ \emph {et~al.}(1995)\citenamefont
  {Farakos}, \citenamefont {Kajantie}, \citenamefont {Rummukainen},\ and\
  \citenamefont {Shaposhnikov}}]{Farakos:1994xh}%
  \BibitemOpen
  \bibfield  {author} {\bibinfo {author} {\bibfnamefont {K.}~\bibnamefont
  {Farakos}}, \bibinfo {author} {\bibfnamefont {K.}~\bibnamefont {Kajantie}},
  \bibinfo {author} {\bibfnamefont {K.}~\bibnamefont {Rummukainen}}, \ and\
  \bibinfo {author} {\bibfnamefont {M.~E.}\ \bibnamefont {Shaposhnikov}},\
  }\href {\doibase 10.1016/0550-3213(95)80129-4} {\bibfield  {journal}
  {\bibinfo  {journal} {Nucl. Phys.}\ }\textbf {\bibinfo {volume} {B442}},\
  \bibinfo {pages} {317} (\bibinfo {year} {1995})},\ \Eprint
  {http://arxiv.org/abs/hep-lat/9412091}{arXiv:hep-lat/9412091
  [hep-lat]}\BibitemShut {NoStop}%
\bibitem [{\citenamefont {Kajantie}\ \emph
  {et~al.}(1996{\natexlab{a}})\citenamefont {Kajantie}, \citenamefont {Laine},
  \citenamefont {Rummukainen},\ and\ \citenamefont
  {Shaposhnikov}}]{Kajantie:1995dw}%
  \BibitemOpen
  \bibfield  {author} {\bibinfo {author} {\bibfnamefont {K.}~\bibnamefont
  {Kajantie}}, \bibinfo {author} {\bibfnamefont {M.}~\bibnamefont {Laine}},
  \bibinfo {author} {\bibfnamefont {K.}~\bibnamefont {Rummukainen}}, \ and\
  \bibinfo {author} {\bibfnamefont {M.~E.}\ \bibnamefont {Shaposhnikov}},\
  }\href {\doibase 10.1016/0550-3213(95)00549-8} {\bibfield  {journal}
  {\bibinfo  {journal} {Nucl. Phys.}\ }\textbf {\bibinfo {volume} {B458}},\
  \bibinfo {pages} {90} (\bibinfo {year} {1996}{\natexlab{a}})},\ \Eprint
  {http://arxiv.org/abs/hep-ph/9508379}{arXiv:hep-ph/9508379
  [hep-ph]}\BibitemShut {NoStop}%
\bibitem [{\citenamefont {Kajantie}\ \emph
  {et~al.}(1996{\natexlab{b}})\citenamefont {Kajantie}, \citenamefont {Laine},
  \citenamefont {Rummukainen},\ and\ \citenamefont
  {Shaposhnikov}}]{Kajantie:1995kf}%
  \BibitemOpen
  \bibfield  {author} {\bibinfo {author} {\bibfnamefont {K.}~\bibnamefont
  {Kajantie}}, \bibinfo {author} {\bibfnamefont {M.}~\bibnamefont {Laine}},
  \bibinfo {author} {\bibfnamefont {K.}~\bibnamefont {Rummukainen}}, \ and\
  \bibinfo {author} {\bibfnamefont {M.~E.}\ \bibnamefont {Shaposhnikov}},\
  }\href {\doibase 10.1016/0550-3213(96)00052-1} {\bibfield  {journal}
  {\bibinfo  {journal} {Nucl. Phys.}\ }\textbf {\bibinfo {volume} {B466}},\
  \bibinfo {pages} {189} (\bibinfo {year} {1996}{\natexlab{b}})},\ \Eprint
  {http://arxiv.org/abs/hep-lat/9510020}{arXiv:hep-lat/9510020
  [hep-lat]}\BibitemShut {NoStop}%
\bibitem [{\citenamefont {Gynther}\ and\ \citenamefont
  {Vepsalainen}(2006)}]{Gynther:2005dj}%
  \BibitemOpen
  \bibfield  {author} {\bibinfo {author} {\bibfnamefont {A.}~\bibnamefont
  {Gynther}}\ and\ \bibinfo {author} {\bibfnamefont {M.}~\bibnamefont
  {Vepsalainen}},\ }\href {\doibase 10.1088/1126-6708/2006/01/060} {\bibfield
  {journal} {\bibinfo  {journal} {JHEP}\ }\textbf {\bibinfo {volume} {01}},\
  \bibinfo {pages} {060} (\bibinfo {year} {2006})},\ \Eprint
  {http://arxiv.org/abs/hep-ph/0510375}{arXiv:hep-ph/0510375
  [hep-ph]}\BibitemShut {NoStop}%
\bibitem [{\citenamefont {Laine}\ and\ \citenamefont
  {Meyer}(2015)}]{Laine:2015kra}%
  \BibitemOpen
  \bibfield  {author} {\bibinfo {author} {\bibfnamefont {M.}~\bibnamefont
  {Laine}}\ and\ \bibinfo {author} {\bibfnamefont {M.}~\bibnamefont {Meyer}},\
  }\href {\doibase 10.1088/1475-7516/2015/07/035} {\bibfield  {journal}
  {\bibinfo  {journal} {JCAP}\ }\textbf {\bibinfo {volume} {1507}},\ \bibinfo
  {pages} {035} (\bibinfo {year} {2015})},\ \Eprint
  {http://arxiv.org/abs/1503.04935}{arXiv:1503.04935 [hep-ph]}\BibitemShut
  {NoStop}%
\bibitem [{\citenamefont {D'Onofrio}\ and\ \citenamefont
  {Rummukainen}(2016)}]{DOnofrio:2015gop}%
  \BibitemOpen
  \bibfield  {author} {\bibinfo {author} {\bibfnamefont {M.}~\bibnamefont
  {D'Onofrio}}\ and\ \bibinfo {author} {\bibfnamefont {K.}~\bibnamefont
  {Rummukainen}},\ }\href {\doibase 10.1103/PhysRevD.93.025003} {\bibfield
  {journal} {\bibinfo  {journal} {Phys. Rev.}\ }\textbf {\bibinfo {volume}
  {D93}},\ \bibinfo {pages} {025003} (\bibinfo {year} {2016})},\ \Eprint
  {http://arxiv.org/abs/1508.07161}{arXiv:1508.07161 [hep-ph]}\BibitemShut
  {NoStop}%
\bibitem [{\citenamefont {Belanger}\ \emph {et~al.}(2018)\citenamefont
  {Belanger}, \citenamefont {Boudjema}, \citenamefont {Goudelis}, \citenamefont
  {Pukhov},\ and\ \citenamefont {Zaldivar}}]{Belanger:2018mqt}%
  \BibitemOpen
  \bibfield  {author} {\bibinfo {author} {\bibfnamefont {G.}~\bibnamefont
  {Belanger}}, \bibinfo {author} {\bibfnamefont {F.}~\bibnamefont {Boudjema}},
  \bibinfo {author} {\bibfnamefont {A.}~\bibnamefont {Goudelis}}, \bibinfo
  {author} {\bibfnamefont {A.}~\bibnamefont {Pukhov}}, \ and\ \bibinfo {author}
  {\bibfnamefont {B.}~\bibnamefont {Zaldivar}},\ }\href {\doibase
  10.1016/j.cpc.2018.04.027} {\bibfield  {journal} {\bibinfo  {journal}
  {Comput. Phys. Commun.}\ }\textbf {\bibinfo {volume} {231}},\ \bibinfo
  {pages} {173} (\bibinfo {year} {2018})},\ \Eprint
  {http://arxiv.org/abs/1801.03509}{arXiv:1801.03509 [hep-ph]}\BibitemShut
  {NoStop}%
\bibitem [{\citenamefont {Gondolo}\ and\ \citenamefont
  {Gelmini}(1991)}]{Gondolo:1990dk}%
  \BibitemOpen
  \bibfield  {author} {\bibinfo {author} {\bibfnamefont {P.}~\bibnamefont
  {Gondolo}}\ and\ \bibinfo {author} {\bibfnamefont {G.}~\bibnamefont
  {Gelmini}},\ }\href {\doibase 10.1016/0550-3213(91)90438-4} {\bibfield
  {journal} {\bibinfo  {journal} {Nucl. Phys.}\ }\textbf {\bibinfo {volume}
  {B360}},\ \bibinfo {pages} {145} (\bibinfo {year} {1991})}\BibitemShut
  {NoStop}%
\bibitem [{\citenamefont {Hindmarsh}\ and\ \citenamefont
  {Philipsen}(2005)}]{Hindmarsh:2005ix}%
  \BibitemOpen
  \bibfield  {author} {\bibinfo {author} {\bibfnamefont {M.}~\bibnamefont
  {Hindmarsh}}\ and\ \bibinfo {author} {\bibfnamefont {O.}~\bibnamefont
  {Philipsen}},\ }\href {\doibase 10.1103/PhysRevD.71.087302} {\bibfield
  {journal} {\bibinfo  {journal} {Phys. Rev.}\ }\textbf {\bibinfo {volume}
  {D71}},\ \bibinfo {pages} {087302} (\bibinfo {year} {2005})},\ \Eprint
  {http://arxiv.org/abs/hep-ph/0501232}{arXiv:hep-ph/0501232
  [hep-ph]}\BibitemShut {NoStop}%
\bibitem [{\citenamefont {Karsch}\ \emph {et~al.}(2000)\citenamefont {Karsch},
  \citenamefont {Laermann},\ and\ \citenamefont {Peikert}}]{Karsch:2000ps}%
  \BibitemOpen
  \bibfield  {author} {\bibinfo {author} {\bibfnamefont {F.}~\bibnamefont
  {Karsch}}, \bibinfo {author} {\bibfnamefont {E.}~\bibnamefont {Laermann}}, \
  and\ \bibinfo {author} {\bibfnamefont {A.}~\bibnamefont {Peikert}},\ }\href
  {\doibase 10.1016/S0370-2693(00)00292-6} {\bibfield  {journal} {\bibinfo
  {journal} {Phys. Lett.}\ }\textbf {\bibinfo {volume} {B478}},\ \bibinfo
  {pages} {447} (\bibinfo {year} {2000})},\ \Eprint
  {http://arxiv.org/abs/hep-lat/0002003}{arXiv:hep-lat/0002003
  [hep-lat]}\BibitemShut {NoStop}%
\bibitem [{\citenamefont {Drees}\ \emph {et~al.}(2015)\citenamefont {Drees},
  \citenamefont {Hajkarim},\ and\ \citenamefont {Schmitz}}]{Drees:2015exa}%
  \BibitemOpen
  \bibfield  {author} {\bibinfo {author} {\bibfnamefont {M.}~\bibnamefont
  {Drees}}, \bibinfo {author} {\bibfnamefont {F.}~\bibnamefont {Hajkarim}}, \
  and\ \bibinfo {author} {\bibfnamefont {E.~R.}\ \bibnamefont {Schmitz}},\
  }\href {\doibase 10.1088/1475-7516/2015/06/025} {\bibfield  {journal}
  {\bibinfo  {journal} {JCAP}\ }\textbf {\bibinfo {volume} {1506}},\ \bibinfo
  {pages} {025} (\bibinfo {year} {2015})},\ \Eprint
  {http://arxiv.org/abs/1503.03513}{arXiv:1503.03513 [hep-ph]}\BibitemShut
  {NoStop}%
\bibitem [{\citenamefont {Bazavov}\ \emph {et~al.}(2014)\citenamefont {Bazavov}
  \emph {et~al.}}]{Bazavov:2014pvz}%
  \BibitemOpen
  \bibfield  {author} {\bibinfo {author} {\bibfnamefont {A.}~\bibnamefont
  {Bazavov}} \emph {et~al.} (\bibinfo {collaboration} {HotQCD}),\ }\href
  {\doibase 10.1103/PhysRevD.90.094503} {\bibfield  {journal} {\bibinfo
  {journal} {Phys. Rev.}\ }\textbf {\bibinfo {volume} {D90}},\ \bibinfo {pages}
  {094503} (\bibinfo {year} {2014})},\ \Eprint
  {http://arxiv.org/abs/1407.6387}{arXiv:1407.6387 [hep-lat]}\BibitemShut
  {NoStop}%
\bibitem [{\citenamefont {Boyd}\ \emph {et~al.}(1996)\citenamefont {Boyd},
  \citenamefont {Engels}, \citenamefont {Karsch}, \citenamefont {Laermann},
  \citenamefont {Legeland}, \citenamefont {Lutgemeier},\ and\ \citenamefont
  {Petersson}}]{Boyd:1996bx}%
  \BibitemOpen
  \bibfield  {author} {\bibinfo {author} {\bibfnamefont {G.}~\bibnamefont
  {Boyd}}, \bibinfo {author} {\bibfnamefont {J.}~\bibnamefont {Engels}},
  \bibinfo {author} {\bibfnamefont {F.}~\bibnamefont {Karsch}}, \bibinfo
  {author} {\bibfnamefont {E.}~\bibnamefont {Laermann}}, \bibinfo {author}
  {\bibfnamefont {C.}~\bibnamefont {Legeland}}, \bibinfo {author}
  {\bibfnamefont {M.}~\bibnamefont {Lutgemeier}}, \ and\ \bibinfo {author}
  {\bibfnamefont {B.}~\bibnamefont {Petersson}},\ }\href {\doibase
  10.1016/0550-3213(96)00170-8} {\bibfield  {journal} {\bibinfo  {journal}
  {Nucl. Phys.}\ }\textbf {\bibinfo {volume} {B469}},\ \bibinfo {pages} {419}
  (\bibinfo {year} {1996})},\ \Eprint
  {http://arxiv.org/abs/hep-lat/9602007}{arXiv:hep-lat/9602007
  [hep-lat]}\BibitemShut {NoStop}%
\bibitem [{\citenamefont {Steigman}\ \emph {et~al.}(2012)\citenamefont
  {Steigman}, \citenamefont {Dasgupta},\ and\ \citenamefont
  {Beacom}}]{Steigman:2012nb}%
  \BibitemOpen
  \bibfield  {author} {\bibinfo {author} {\bibfnamefont {G.}~\bibnamefont
  {Steigman}}, \bibinfo {author} {\bibfnamefont {B.}~\bibnamefont {Dasgupta}},
  \ and\ \bibinfo {author} {\bibfnamefont {J.~F.}\ \bibnamefont {Beacom}},\
  }\href {\doibase 10.1103/PhysRevD.86.023506} {\bibfield  {journal} {\bibinfo
  {journal} {Phys. Rev. D}\ }\textbf {\bibinfo {volume} {86}},\ \bibinfo
  {pages} {023506} (\bibinfo {year} {2012})},\ \Eprint
  {http://arxiv.org/abs/1204.3622}{arXiv:1204.3622 [hep-ph]}\BibitemShut
  {NoStop}%
\bibitem [{\citenamefont {Komatsu}\ \emph {et~al.}(2011)\citenamefont {Komatsu}
  \emph {et~al.}}]{Komatsu:2010fb}%
  \BibitemOpen
  \bibfield  {author} {\bibinfo {author} {\bibfnamefont {E.}~\bibnamefont
  {Komatsu}} \emph {et~al.} (\bibinfo {collaboration} {WMAP}),\ }\href
  {\doibase 10.1088/0067-0049/192/2/18} {\bibfield  {journal} {\bibinfo
  {journal} {Astrophys. J. Suppl.}\ }\textbf {\bibinfo {volume} {192}},\
  \bibinfo {pages} {18} (\bibinfo {year} {2011})},\ \Eprint
  {http://arxiv.org/abs/1001.4538}{arXiv:1001.4538 [astro-ph.CO]}\BibitemShut
  {NoStop}%
\end{thebibliography}%

\end{document}